\documentclass{aa}

\usepackage[utf8]{inputenc}
\usepackage[cdot,squaren]{SIunits}
\usepackage{numprint}
\usepackage[T1]{fontenc}
\usepackage[varg]{txfonts}
\usepackage{amssymb}
\usepackage{graphicx}
\usepackage{tabularx} 
\usepackage{mathrsfs}
\usepackage{subcaption}
\usepackage[version=4]{mhchem}
\usepackage{tabularx}
\usepackage{multicol}
\usepackage{listings}
\usepackage{tikz}
\usetikzlibrary{patterns,patterns.meta}
\everymath{\displaystyle}
\usepackage{ulem}

\newcommand\Stokes{\mbox{St}}  

\begin{document}
\title{Planetesimal gravitational collapse in gaseous environment: thermal and dynamic evolution}

\author{P. Segretain \inst{\ref{inst1}} \and H. Méheut \inst{\ref{inst1}} \and M. Moreira \inst{\ref{inst2}} \and G. Lesur \inst{\ref{inst3}} \and C. Robert \inst{\ref{inst3}} \and J. Mauxion \inst{\ref{inst3}}}

\institute{Université Côte d'Azur, Observatoire de la Côte d'Azur, CNRS, Laboratoire Lagrange, Bd de l'Observatoire, CS 34229, 06304 Nice cedex 4, France \email{paul.segretain@oca.eu} \label{inst1} \and ISTO, Institut des Sciences de la Terre d'Orléans, 45071 Orléans, France \label{inst2} \and Univ. Grenoble Alpes, CNRS, IPAG, 38000 Grenoble, France \label{inst3}}

\date{Received XX \ Accepted YY}

\abstract{Planetesimal formation models often invoke gravitational collapse of pebble clouds to overcome various barriers to grain growth and propose processes to concentrate particles sufficiently to trigger this collapse. On the other hand, the geochemical approach to planet formation constrains the conditions for planetesimal formation and evolution by providing temperatures that should be reached to explain the final composition of planetesimals, the building blocks of planets. To elucidate the thermal evolution during gravitational collapse, we use numerical simulations of a self-gravitating cloud of particles and gas coupled by gas drag. Our goal is to determine how the gravitational energy relaxed during the contraction is distributed among the different energy components of the system, and how this constrains the thermal and dynamical planetesimals' history. We identify the conditions necessary to achieve a temperature increase of several hundred Kelvins, potentially reaching up to 1600 K. Our results emphasize the key role of the gas during the collapse.}


\titlerunning{Thermodynamics of planetesimal collapse}
\authorrunning{P. Segretain et al.}

\maketitle

\section{Introduction}
The formation of planetesimals is one of the key open questions in planet formation. Not only is it necessary to identify the properties of the building blocks of planets but determining the physical conditions experienced during the formation of parent bodies could also benefit the understanding of the chemical and isotopic composition of small bodies of the Solar System and meteorites \citep{boyet_enstatite_2018,zhu_chondrite_2023}. These gravitationally bound objects have a size of $1-\unit{1000}{\kilo\meter}$ and a common view is that their destructive collision cascade is at the origin of the asteroid and Kuiper belts as well as the Oort cloud in the Solar System \citep{morbidelli_chapter_2020}, whereas protoplanets are the result of their constructive collisions and growth \citep{johansen_forming_2017}.
Indeed, both the isotopic chronology of meteorites, which are fragments of asteroids and the detection of protoplanets within gaseous protoplanetary discs \citep{keppler_discovery_2018,gratton_blobs_2019} suggest that planetesimals form during the first Myr of the stellar system, in the gaseous protoplanetary disc era \citep{kruijer_protracted_2014}. 

However, the pebble-sized bodies from which planetesimals arise drift radially through the disc towards the star on very short timescales of hundreds to thousands of years due to the gas drag \citep{weidenschilling_aerodynamics_1977} while the bouncing, erosion or fragmentation barriers could severely limit their growth \citep{blum_dust_2018}. Recently, simulations of dust aggregates of sub-micrometre monomers show that their growth by collisions could allow to overcome the fragmentation barrier. \citep[see e.g.][]{hasegawa_collisional_2021,hasegawa_collisional_2023}. In this model \citet{kobayashi_rapid_2021,kobayashi_rapid_2023} show that pebbles could form planetesimals and then gas giant planets in a relatively short time. Another path to overcome these barriers and form planetesimals commonly considered invokes a strong concentration of grains, followed by a gravitational collapse on a short timescale. The physical processes responsible for the strong clustering of solids are currently under debate, including in particular the streaming instability \citep[e.g.][]{youdin_streaming_2005, carrera_streaming_2022}, the concentration in magneto-rotational instability zonal flows \citep[e.g.][]{fromang_global_2006, johansen_high-resolution_2011, dittrich_gravoturbulent_2013} but also instabilities forming vortices such as the Rossby wave instability \citep[e.g.][]{lovelace_rossby_1999, meheut_dust-trapping_2012}, the vertical shear instability \citep[e.g.][]{urpin_magnetic_1998, nelson_linear_2013, manger_vortex_2018} or the convective over-stability \citep[e.g.][]{klahr_convective_2014,lyra_convective_2014,raettig_pebble_2021}, as well as turbulent clustering \citep[e.g.][]{cuzzi_size-selective_2001, pan_turbulent_2011,gerosa_clusters_2023}. In most of the works cited above, only one or a small number of sizes were considered for the solids. The efficiency of these different mechanisms when considering a realistic size distribution is a point of debate. For example, in the case of streaming instability, \citet{krapp_streaming_2019} claimed that including a particle size distribution could greatly reduce the instability growth time, while \citet{schaffer_streaming_2021} found that dust size distribution changes the dynamics but does not suppress strong clustering. All of these clustering processes rely on a balanced coupling between the gas and the solid phase: the solid has to follow but only partially the gas dynamics, and possibly modify this gas dynamics as in the streaming instability. This means that the solid Stokes number generally lies in the range $[10^{-2}; 1]$ where the Stokes number \Stokes $\ $quantifies the ratio between the stopping time, the typical timescale required for the solid velocity to reach the gas velocity, and the typical timescale of the gas dynamics, which is generally the orbital period \citep[see e.g.][]{lesur_hydro-_2022}.

The following step, the gravitational collapse, is key to characterizing the properties of planetesimals and has been investigated in several ways. The results of collisions on the final grain sizes in the planetesimal were investigated by \citet{jansson_role_2017} and how the initial angular momentum of the particle cloud is responsible for the binary nature of planetesimals by \citet{nesvorny_binary_2021}. More recently, \cite{lorek_formation_2024} evaluated how the initial rotation of a pebble cloud modifies the final shape of the planetesimal formed. All of these studies focused on the dynamics of the solids without considering the gas. This is consistent with the high dust-to-gas ratio and the location considered, in the outer solar system, where the gas density is lower and the stopping time is therefore longer than the collision time. In fact, as the particles that tend to cluster have Stokes number in the range $[10^{-2}; 1]$, the coupling with the gas may also have non-negligible effects during the collapse. \citet{polak_high-resolution_2023} built on the work by \citet{nesvorny_binary_2021} and studied the impact of the initial location of the collapsing clump on the number and size distribution of the formed planetesimals. \citet{wahlberg_jansson_radially_2017} added the gas drag in the dynamics of solids and showed how it modifies the structure of the formed object, with medium-sized pebbles at the centre surrounded by a mixture of large and small pebbles. \citet{shariff_spherically_2015} examined the two-way coupling between gas and solids during the contraction and observed an oscillatory behaviour of the pebble clump. As they point out, for a minimum-mass solar nebula given by \cite{hayashi_structure_1981}, this oscillatory behaviour requires very large and dense clumps, for which no formation process is known. Moreover, these results focus on the isothermal case, where no gas heating process is included. To pave the way towards an understanding of planetesimal formation that is consistent with both the disc gas-dust dynamics and the cosmochemical constraints from the Solar System, we consider here the gas dynamical and thermal evolution during the collapse. As we show below, this approach is particularly important for AU-scale planetesimal formation.

In this paper we consider a gravitational collapse of pebbles, modifying the gas dynamics and temperature. We build on the work of \cite{shariff_spherically_2015}, considering a spherically symmetric collapse, where we take into account the gas thermodynamics and include the frictional heating of the dust on the gas. The paper is structured as follows. In section \ref{model}, we discuss paths to model the collapse of a dust clump coupled to a gaseous environment and introduce some necessary physical concepts. The numerical methods are presented in section \ref{method}, and we discuss the results and their limits in section \ref{discussion}.

\section{Model for pebbles collapse in a fluid}\label{model}

In this study, we assume a spherical pebble cloud collapsing under self-gravity in a gaseous environment. Given the length scales that will be involved in this work, small compared to disc pressure length scales, the gas Keplerian shear and disc rotation are neglected. The exact background gas velocity profile varies with the particle clustering process (in a vortex, in a pressure bump, \ldots), and we here consider the simple case with no initial velocity in the referential centred on the pebble cloud centre. In this section, we present some characteristic times, and the governing equations for gas and solid evolution, before giving a first simple analytical estimate of the gas temperature increase during the collapse.

\subsection{Characteristic times}\label{times}

The gravitational collapse of a coupled pebble and gas cloud involves several timescales.  
The disc dynamical timescale is the Keplerian orbital time, $t_\mathrm{orb}=2\pi \sqrt{{a^3}/{\mathcal{G}M_\odot}}$ where $\mathcal{G}$ is the gravitational constant, $a$ the semi-major axis of the orbit and $M_\odot$ the central star mass. We consider in this work gravitational collapses occurring in the inner disc typically at 1~AU (or less) from the central star. At such a distance, around a solar-type star, $t_\mathrm{orb}\sim 1~\mathrm{yr}$.

We study a spherical clump with a mass $M$ and initial radius $R_0$ composed of gas and particles. The position of a particle located at the cloud surface evolves under the clump gravity and the gas drag which is expected to be linear in the inner part of discs and follows the equation
\begin{equation}
\dfrac{\mathrm{d}^2 r}{\mathrm{d}t^2}=-\dfrac{\mathcal{G}M}{r^2}-\frac{1}{\tau_s}\frac{\mathrm{d}r}{\mathrm{d}t},
\label{eq_diff_fric}
\end{equation}
with $\tau_s $ the particle friction time or stopping time, and where the gas is assumed static.  
To obtain a first rough estimation, we consider the terminal velocity approximation. In this approximation ($\mathrm{d}^2_t r\sim 0$), eq. \ref{eq_diff_fric} gives
\begin{equation}
r(t)=\sqrt[3]{R_0^3-3\tau_s\mathcal{G}Mt}.
\label{radius_friction}
\end{equation}
and the collapse time (when $r=0$) is
\begin{equation}
t_\mathrm{col}=\dfrac{R_0^3}{3\mathcal{G}M\tau_s}=\dfrac{8 t_\mathrm{ff}^2}{3\pi^2\tau_s} = \frac{8 t_\mathrm{ff}}{3\pi^2\Stokes_\mathrm{ff}}\label{col_time},
\end{equation}
with the free-fall time defined as 
\begin{equation*}
    t_\mathrm{ff}=\dfrac{\pi}{2}\sqrt{\dfrac{R_0^3}{2\mathcal{G}M}}
\end{equation*} . We have introduced the Stokes number for this dynamical evolution $\Stokes_\mathrm{ff}~=~\tau_s / t_\mathrm{ff}$. This differs from the standard Stokes number used for protoplanetary discs $\Stokes_\Omega~=~\tau_s\Omega~=~2\pi\tau_s/t_\mathrm{orb}$. Equation \ref{col_time} shows in particular that the pebble collapse time is inversely proportional to the Stokes number $\Stokes_\mathrm{ff}$.

\begin{figure}
\centering
\includegraphics[width=\columnwidth]{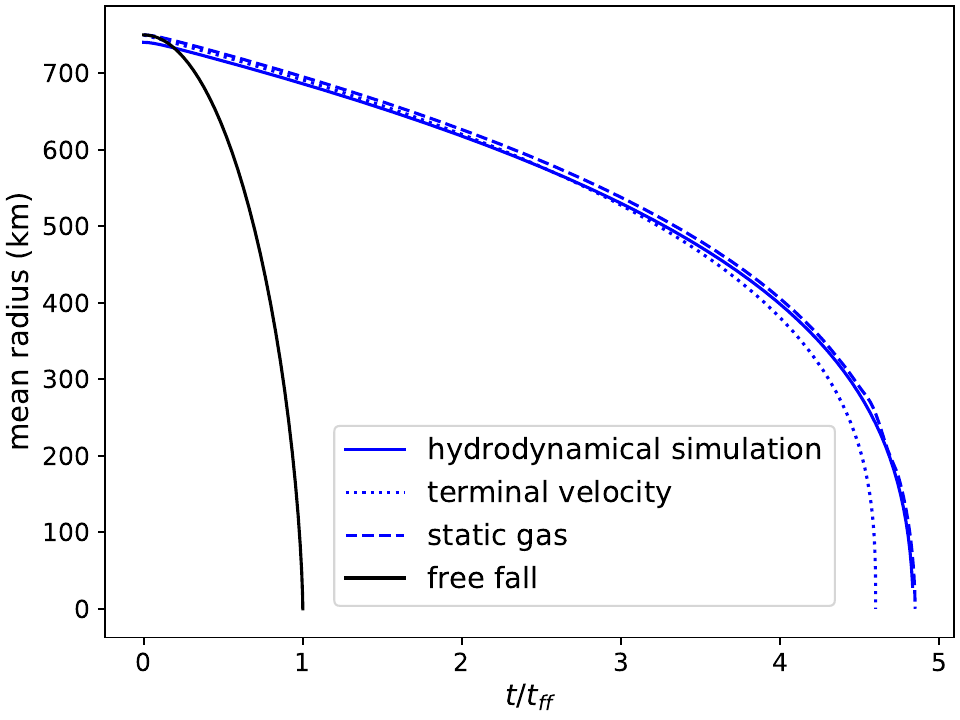}
\caption{Time evolution of the pebble cloud size in the non-linear numerical simulation, a static gas solution (Eq. \ref{eq_diff_fric}), and the analytical solution \ref{radius_friction} in the terminal velocity approximation (respectively solid, dashed and dotted line). The black line corresponds to the standard free fall, without friction (see appendix \ref{ff_time}).}
\label{size_evol}
\end{figure}

In Fig. \ref{size_evol} the time evolution of the clump size estimated in the terminal velocity approximation  (equation \ref{radius_friction}) is compared with the numerical solution of equation \ref{eq_diff_fric} with a static gas, and with a hydrodynamical simulation (fiducial case, see section \ref{init}). The cloud size is characterized by its mean radius, defined as in \citep{shariff_spherically_2015} 
\begin{equation*}
    r_m=\dfrac{\int_0^{+\infty}r\rho_\mathrm{p}(r)\mathrm{d}r}{\int_0^{+\infty}\rho_\mathrm{p}(r)\mathrm{d}r} .
\end{equation*}
 For a filled sphere of uniform density and radius $R_0$, the value of $r_m$ is $0.5R_0$. Figure \ref{size_evol} shows that the terminal velocity approximation (eq. \ref{col_time}) underestimates by about 10\% the collapse time when compared with hydrodynamical simulations. $t_\mathrm{col}$ (here about \unit{6.10^6}{\second}) is, therefore, a good proxy for the collapse time. Moreover, the static gas solution matches very well with the numerical simulation. It thus provides an upper limit for the simulation integration time.

\subsection{Governing equations}\label{gov_eq}

The gas dynamics is given by the inviscid Euler equations
\begin{align}
  {\partial_t \rho_\mathrm{g}} + \boldsymbol{\nabla}\cdot(\rho_g \boldsymbol{u})= {}& 0 \label{Euler_rho}\\ 
  {\partial_t}\rho_\mathrm{g} \boldsymbol{u} + \boldsymbol{\nabla}\cdot(\rho_\mathrm{g} \boldsymbol{u} \otimes \boldsymbol{u}+P)={}& -\rho_\mathrm{g}\boldsymbol{\nabla}(\phi_\mathrm{g+p}+\phi_c) - \rho_\mathrm{p} \boldsymbol{a}_{\mathrm{drag}}\label{Euler_u}\\
{\partial_t E}+ \boldsymbol{\nabla}\cdot\left( \boldsymbol{u} (E+P)\right) = {}& -\rho_\mathrm{g} \boldsymbol{u} \cdot \boldsymbol{\nabla}(\phi_\mathrm{g+p}+\phi_c) + \mathcal{P}_{\mathrm{drag}}\label{Euler_e}
  \end{align}
where $\rho_\mathrm{g}$, $\boldsymbol{u}$, $P$ and $E$ are respectively the mass density, the velocity, the pressure and the total energy (kinetic and internal) of the gas. We consider the self-gravity of the gas and dust, and the drag force between them. In the momentum equation, $\rho_\mathrm{p}$ is the local pebble mass density. The equation system is closed with an ideal equation of states written as
\begin{equation}
    P =\rho_\mathrm{g} \dfrac{R}{M}T,\ U = \dfrac{P}{\gamma-1}
\end{equation}
where  $R$, $M$, $T$, $U$ and $\gamma$ are the ideal gas constant, the molar mass, the temperature, the internal energy and the adiabatic index of the gas.

Pebbles are modelled as a pressure-less fluid whose dynamic is given by 
\begin{align}
  {\partial_t \rho_\mathrm{d}} + \boldsymbol{\nabla}\cdot(\rho_d \boldsymbol{v_\mathrm{d}})= {}& 0 \label{Euler_rho_dust}\\ 
  {\partial_t}\rho_\mathrm{d} \boldsymbol{v_\mathrm{d}} + \boldsymbol{\nabla}\cdot(\rho_\mathrm{d} \boldsymbol{v_\mathrm{d}} \otimes \boldsymbol{v_\mathrm{d}})={}& -\rho_\mathrm{d}\boldsymbol{\nabla}(\phi_\mathrm{g+p}+\phi_c) + \rho_\mathrm{d} \boldsymbol{a}_{\mathrm{drag}}\label{Euler_u_dust}
  \end{align}
where $\rho_\mathrm{d}$ and $\boldsymbol{v_\mathrm{d}}$ are the mass density and the velocity of the fluid.
$\phi_\mathrm{g+p}$ is the gravitational potential of both gas and pebbles, which solves the Poisson equation
\begin{equation}
    \Delta \phi_\mathrm{g+p} = 4\pi \mathcal{G} \rho_\mathrm{g+p}
\end{equation}
where $\rho_\mathrm{g+p}$ is the total mass density of gas and pebbles. $\phi_c$ is addressed in section \ref{method} where its introduction is justified by the limited simulation domain.

In these equations, $\boldsymbol{a}_{\mathrm{drag}}$ is the drag acceleration exerted by the gas on pebbles (and $-\boldsymbol{a}_{\mathrm{drag}}$ the feedback onto the gas). The drag acceleration is computed with the prescription 
\begin{equation}
    \boldsymbol{a}_{\mathrm{drag}} = -\dfrac{\boldsymbol{v_\mathrm{d}} - \boldsymbol{u}}{\tau_s}
\end{equation}
where $\tau_s$ is the dust stopping time given by
\begin{equation}
    \tau_s = 
    \begin{cases}
        \dfrac{\rho_m a}{\rho_\mathrm{g} \overline{c}} \text{ if } a < {9\lambda}/{4} \text{ \citep[Epstein regime,][]{epstein_resistance_1924}} \\
        \dfrac{4\rho_m a^2}{9\rho_\mathrm{g} \overline{c}\lambda} \text{ else (Stokes regime)}
    \end{cases}
    \label{def_tau}
\end{equation}
depending on the material density $\rho_m$ and size $a$ of the pebbles, the density $\rho_\mathrm{g}$ of the gas and the mean thermal speed of the gas molecules $\overline{c}=\sqrt{{8}/{\pi}}c_s$, with $c_s$ the sound speed in the gas. $\lambda$ is the mean-free path of the gas molecules. In the energy equation, $\mathcal{P}_{\mathrm{drag}}$ is the power given by the particles to the gas, defined as
\begin{equation}
    \mathcal{P}_{\mathrm{drag}} = \rho_\mathrm{p} \dfrac{\boldsymbol{v_\mathrm{p}} - \boldsymbol{u}}{\tau_s}\cdot \boldsymbol{u} + \rho_\mathrm{p} \dfrac{(\boldsymbol{v_\mathrm{p}} - \boldsymbol{u})^2}{\tau_s}.
    \label{edrag}
\end{equation}
This includes the power done by the drag force and the frictional heating which can be considered as irreversible power generating entropy.

Finally, in the collapsing area, the pebbles are highly concentrated compared to the disc mean solid density, it can be considered as regions with a large concentration of dust of many sizes (not included in the simulations) and therefore optically thick. We then consider an adiabatic evolution for the system of gas and pebbles.

\subsection{Static gas}\label{static_gas}

An estimation of the heating by gravitational collapse could be provided assuming here a static gas and using an energetic approach where the initial gravitational energy of the pebbles would be fully converted into gas internal energy. This would mean that the pebbles fully transfer their kinetic energy to gas through friction. This simple approach mimics some estimations made to obtain an upper limit on the temperature of a geophysical body \citep[see e.g.][section 2.1.1]{lichtenberg_geophysical_2022}.

The initial condition is a spherical pebble cloud with a uniform mass density\footnote{We define here a \textit{ball} as a filled sphere or a solid sphere, and a \textit{sphere} as the surface of a ball}. The gravitational energy of the ball of mass $M$ and initial radius $R_0$, is $E_\mathcal{G} = -{3}{\mathcal{G}M^2}/{5}{R_0}$. An infinitesimal contraction of the pebble ball releases the gravitational energy 
\begin{equation}
    \mathrm{d}E_\mathcal{G} = \dfrac{3}{5}\dfrac{\mathcal{G}M^2}{R^2}\mathrm{d}R.
\end{equation}
Assuming this released gravitational energy spread uniformly in the ball of the same radius $R$, the gas temperature would increase by this infinitesimal contraction by
\begin{equation}
    \mathrm{d} T = \left(\dfrac{c_v}{V_m} \dfrac{4}{3}\pi R^3\right)^{-1}\mathrm{d}E_\mathrm{p}= \dfrac{9\mathcal{G}M^2}{20\pi c_v/V_m}\dfrac{\mathrm{d}R}{R^5},
\end{equation}
with $c_v$ the molar heat capacity at constant volume and $V_m$ the molar volume. The gas shell of radius $\Tilde{R}$ receive energy from the solids through drag only when the pebble cloud is larger than $\Tilde{R}$. The temperature at a radius $\Tilde{R}$  would have increased by the end of the collapse of
\begin{equation}
    \Delta T(\Tilde{R}) = \int_{\Tilde{R}}^{R_0} \dfrac{\mathrm{d}\Delta T}{\mathrm{d}r} \mathrm{d}r = \dfrac{9\mathcal{G}M^2}{20\pi c_v/V_m}\dfrac{1}{4}\left(\dfrac{1}{\Tilde{R}^4}-\dfrac{1}{R_0^4}\right).
    \label{delta_temp}
\end{equation}
For a pebble cloud of mass $\unit{3\times10^{16}}{\kilogram}$ contracting from $R_0=\unit{1500}{\kilo\meter}$ to $\Tilde{R}=\unit{50}{\kilo\meter}$, the gravitational energy released is $\Delta E_\mathcal{G} = 3\mathcal{G}M^2/5 \times(1/\Tilde{R}-1/R_0) \sim \unit{7\times 10^{17}}{\joule}$ . If this energy is deposited through the process described above in a gas initially at $\unit{320}{\kelvin}$ and $\unit{8.34}{\pascal}$ composed fully of \ce{H2}, (which corresponds to a total mass of about $\unit{9\times\power{10}{13}}{\kilogram}$, and $c_v=\unit{20.8}{\joule\usk\reciprocal\mole\usk\reciprocal\kelvin}$) the temperature would reach more than \unit{5000}{\kelvin} according to equation \ref{delta_temp}. 

This simple approach gives an upper limit to gas heating. The very high temperature obtained here shows that the hypothesis of a full conversion of gravitational energy into gas internal energy is not correct.  Therefore a more complete approach, including the gas and dust thermal as well as dynamical evolution with fully non-linear numerical simulations is necessary.

\section{Numerical methods}\label{method}

We perform 1D spherically symmetric simulations using the \texttt{IDEFIX} code \citep{lesur_idefix_2023}, solving the hydrodynamical equations with finite-volume methods through a Godunov scheme. \texttt{IDEFIX} also includes a self-gravity module \citep[appendix A]{mauxion_modeling_2024} to solve the Poisson equation with linear algebra methods. The solids can be modelled in two ways. In this paper, we choose to model them as a pressure-less fluid. We provide a comparison with a Lagrangian particle model in appendix \ref{LP}, showing that for our problem the two approaches are similar. The runs have been performed using a second-order Runge-Kutta scheme, second-order spatial reconstruction and the Lax-Friedrichs Riemann solver.

\subsection{Initial conditions}\label{init}

Initially, the pebbles are static, forming a clump with a uniform density profile in the core and a Gaussian outer profile decrease: 
\begin{equation}
\rho_\mathrm{p}(r) = \begin{cases}
\rho_c \text{ if } r \leq r_c \\
\rho_c \exp\left(-\dfrac{(r-r_c)^2}{\sigma^2}\right) \text{ elsewhere } 
\end{cases}
\label{profile}
\end{equation}
$r_c$ is the initial core radius of the clump, and we define the initial radius of the pebble sphere as $R_0=r_c+\sigma$ with $\sigma = {r_c}/{5}$. The gas has no initial velocity, a uniform density $\rho_0$ and temperature $T_g$  with an ideal equation of state. The adiabatic index of the gas is 1.4. The initial stopping time $\tau_\mathrm{p}$ is $\unit{8.4\times\power{10}{4}}{\second}$ corresponding (for the chosen gas density and sound speed) to pebbles with $a=\unit{15}{\centi\meter}$ and $\rho_m=\unit{3}{\gram\usk\centi\meter\rpcubed}$, and to $\Stokes_\mathrm{ff}=0.06$. For these pebbles, the stopping time is computed in the Stokes regime.

Our fiducial simulation have a pebble mass $M_\mathrm{fid}$ of $\unit{3\mathbf{\times}\power{10}{16}}{\kilogram}$ and gas density $\rho_\mathrm{fid}$ of $\unit{7\mathbf{\times}\power{10}{-6}}{\kilogram\usk\meter\rpcubed}$. This fiducial mass corresponds to a core density $\rho_c$ of $\unit{2.1\times\power{10}{-3}}{\kilogram\usk\meter\rpcubed}$ in equation\ref{profile}, and thus to an initial dust-to-gas ratio of 300.
We explore how the total pebble mass $M_{tot}$ and initial gas density $\rho_g$ modify the final temperature. The values employed for the various runs are presented in table \ref{tab:values}.
\begin{table}[bp]
    \centering
    \begin{tabular}{cc}
        Parameter & Value \\
        \hline
        $M_{tot}$ &  $\lbrace 1;1.7;2.3;3.3;5\rbrace\ M_\mathrm{fid}$\\
        $\rho_g$ & $\lbrace 0.5;0.7;1;1.4;2.1;3\rbrace\ \rho_\mathrm{fid}$ \\
        $a$ & $\unit{\lbrace 5;10;15;20\rbrace}{\centi\meter}$ \\
        $\rho_m$ & $\unit{\lbrace 1;3;5;7\rbrace}{\gram\usk\centi\meter\rpcubed}$ \\
    \end{tabular}
    \caption{Range of values used for our runs.}
    \label{tab:values}
\end{table}
Assuming a planetesimal density of \unit{2000}{\kilogram\usk\meter\rpcubed} (which would correspond to planetesimals made of silicates with some porosity), these initial masses correspond to spherical planetesimals with radii between 10.6 and \unit{26.1}{\kilo\meter}. The initial size and mass of our pebble cloud are therefore small compared to the characteristic size of the parent bodies of the current main belt asteroids. The initial gas temperature ($\unit{320}{\kelvin}$) corresponds to the inner region of a solar-type star protoplanetary disc.  
The initial pebble clump radius $R_0$ is  $\unit{1500}{\kilo\meter}$ (that is $r_c=\unit{1250}{\kilo\meter}$ and $\sigma=\unit{250}{\kilo\meter}$). 

\subsection{Grid and resolution}

We perform 1D spherically symmetric simulations with a grid extending from $R_{in}=0.0013 R_0$ to $R_{out}=4 R_0$ or equivalently $[2, 6000]$ km. The grid is uniform from ${2}$ to $\unit{30}{\kilo\meter}$ and logarithmic from ${30}$ to $\unit{6000}{\kilo\meter}$ giving a good balance between precision and numerical time. A grid comprising 2048 cells is considered, of which 306 are located in the uniform part from 2 to \unit{30}{\kilo\meter} resulting in cells of \unit{90}{\meter}. In the logarithmic part, the ratio between the grid spacing and position $\Delta x/x $ is approximately $0.003$. This distribution of cells is designed such that the cell size in the uniform zone is comparable to the size of the first cell in the logarithmic zone. In order to circumvent the singularity at the centre and the considerable computational expense, the central zone between 0 and $\unit{2}{\kilo\meter}$ is not included in the simulation. Nevertheless, the mass of the gas and particles situated within this zone is represented by a central point mass, with the gravitational potential, designated as $\phi_c$, being incorporated into the self-gravity potential in equations \ref{Euler_u}, \ref{Euler_e} and \ref{Euler_u_dust}.

\subsection{Boundary conditions}

At the outer edge, the outflow boundary condition is used with extrapolated primitive quantities. We performed tests with different boundary conditions and no apparent difference has been seen with zero gradient boundary conditions, showing that the size of the grid is chosen large enough to avoid such numerical artefacts. 

At the inner edge, we use a no-inflow boundary condition. The density and pressure (for the gas) are copied from the innermost cell to the ghost zone (no gradient), the radial velocity is capped at 0 in ghost cells, and the mass flux is set to 0 in case of a positive flux entering the simulation grid. The mass lost at the inner edge is appended to the central point mass, thereby contributing to the overall dynamics of the system via the potential $\phi_c$.

\subsection{Test}

To test the code and verify the compatibility of the different modules, we reproduced the results obtained by \cite{shariff_spherically_2015}. The main difference in the approach is the numerical methods used to solve the hydrodynamic equations, and the boundary condition at the inner edge of the grid. Our approach shows good agreement with their results (as presented in appendix \ref{comparison}), proving that we correctly capture the physics. 

\section{Numerical results of gravitational collapses}

\subsection{Fiducial run} 
\begin{figure*}
\centering
\includegraphics[width=\textwidth]{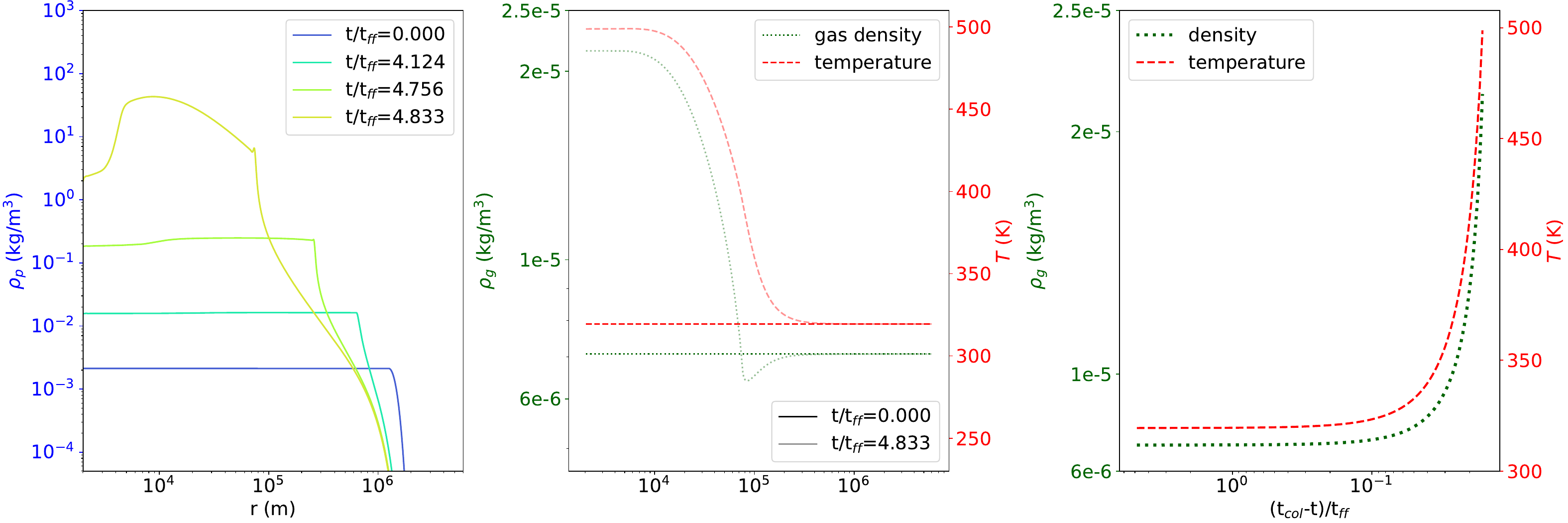}
\caption{(Left) Time evolution of the radial profile of the particle cloud density. (Centre) Gas temperature profiles (dashed) and gas density profiles (dotted), plotted at the beginning and the end of the fiducial run. (Right) Time evolution of the central density (dotted) and temperature (dashed, right axis) of the gas.}

\label{profiles}
\end{figure*}

Figure \ref{profiles} shows radial profiles of the particle density and the evolution of the central gas density and temperature of our fiducial run. This reference run done with the Eulerian approach for the pebbles has a free fall time of $t_\mathrm{ff}=\unit{1.44\times10^6}{\second}$. The coloured curves are traced for different times, that are characterized by the mean radius $r_m$ of the pebble clump. This mean radius evolves from $0.5R_0$ at the beginning of the simulation to $0.02R_0$. This final mean radius corresponds to a core radius of \unit{100}{\kilo\meter}. The collision time starts to be smaller than the friction time (see appendix \ref{comp_col_stop} for estimation of these times). The collisions not being included here, we do not follow further the collapse. The temperature plotted here (red dashed curves) is computed from the primitive variables, density (green dotted curves) and pressure, using the perfect gas law.

The radial density profile of the pebble cloud is shown on the left of Figure \ref{profiles}. The core of the cloud has a radially uniform density for most of the collapse but the concavity of the tail changes and the edge of the cloud becomes steep. Its value increases with time during the collapse, to reach $\unit{30}{\kilogram\usk\meter\rpcubed}$. However, this increase is not regular in time. The initial density is multiplied by 100 before $4.7 t_\mathrm{ff}$, and then multiplied by another factor of 300 at the end of the run, only $0.1 t_\mathrm{ff}$ later. We find the same behaviour for the gas density and temperature shown on the right plot. They are almost constant during the collapse up to $4.7 t_\mathrm{ff}$ (i.e. $(t_\mathrm{col}-t)/t_\mathrm{ff} \sim 10^{-1}$) and change only at the end of the collapse, with a final increase of the central density by a factor 3 and of the gas central temperature by about \unit{170}{\kelvin}. The dashed red curves in the left panel correspond to the temperature profile at the beginning and the end of the collapse. It can be seen that the temperature evolution occurs only within the pebble cloud. In particular, the edge of the temperature rise zone and the edge of the pebble cloud overlap quite well. The increase in gas temperature of \unit{170}{\kelvin}, small compared to the value estimated in section \ref{static_gas}, also shows that the simple approach used there is not correct, as it assumes the pebble kinetic energy to be fully converted into gas heating. In fact, the kinetic energy represents indeed more than half of the released gravitational energy.

\subsection{Heating processes}

\begin{figure}
 \centering
 \includegraphics[width=\columnwidth]{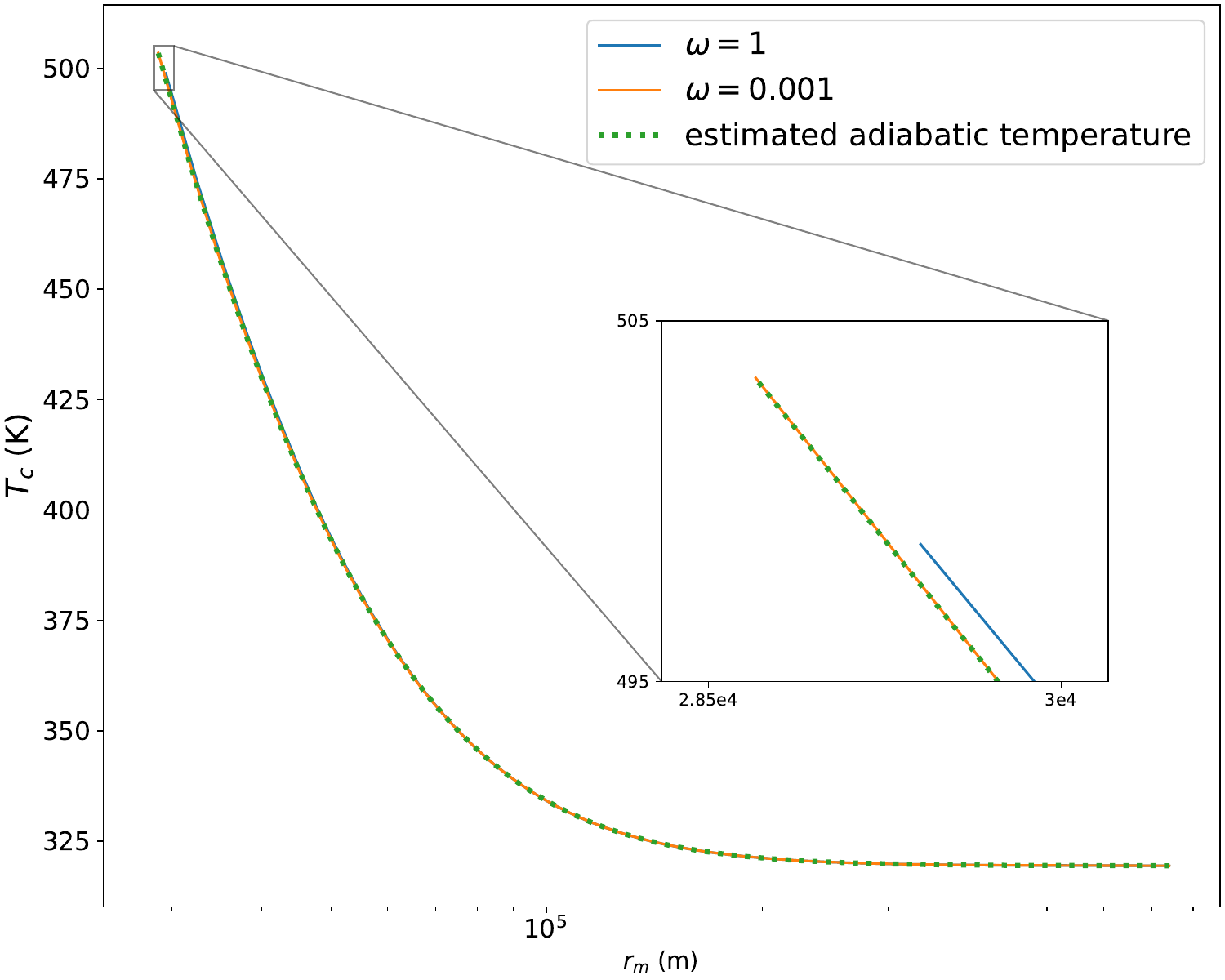}
 \caption{Gas central temperature as a function of the cloud mean radius for two extreme values of $\omega$. The dotted curve corresponds to the theoretical temperature for an adiabatic evolution.}
  \label{compar_om}
\end{figure}

This section examines the heating that occurs during the fiducial run. Initially, the gas is a reservoir of gravitational energy with no kinetic energy, with two external sources of energy: the frictional heating of particles depositing their gravitational energy in the gas, and an energy transfer at the domain outer boundary. The latter would result in a temperature increase also at the outer edge of the computational domain, which we do not see. The central heating is not an outer boundary effect. Therefore, to identify the origin of the gas heating in the fiducial run, we follow \cite{huang_multifluid_2022} and introduce a gas frictional heating parameter $\omega$ in equation \ref{edrag}
\begin{equation}
    \mathcal{P}_{\mathrm{drag}} = \rho_\mathrm{p} \dfrac{\boldsymbol{v_\mathrm{p}} - \boldsymbol{u}}{\tau_s}\cdot \boldsymbol{u} + \omega\rho_\mathrm{p} \dfrac{(\boldsymbol{v_\mathrm{p}} - \boldsymbol{u})^2}{\tau_s}.
    \label{edrag_bis}
\end{equation}
$\omega$ ranges from $0$ for no frictional heating of the gas (in other words this energy would heat the particles) to $\omega=1$ corresponding to a full transfer of the energy dissipated through friction to gas thermal energy.
We compare in Figure \ref{compar_om} the evolution of the central gas temperature during the contraction for two extreme cases $\omega=1$ and $\omega=10^{-3}$.
The weak difference indicates that frictional heating is negligible compared to compressional heating, with a contribution of \unit{1}{\kelvin} to the total temperature increase of about \unit{170}{\kelvin}, as shown by the inset on Figure \ref{compar_om}. As it will be discussed subsequently, this variation is of a similar magnitude to that introduced by modifying the resolution of our grid. Therefore, this is not a physically significant phenomenon. This shows that the heating of the gas is not due to the irreversible term of equation \ref{edrag}. We also note that the density minimum seen in Figure \ref{profiles} is not present for weak frictional heating.

\begin{figure}
 \centering
 \includegraphics[width=\columnwidth]{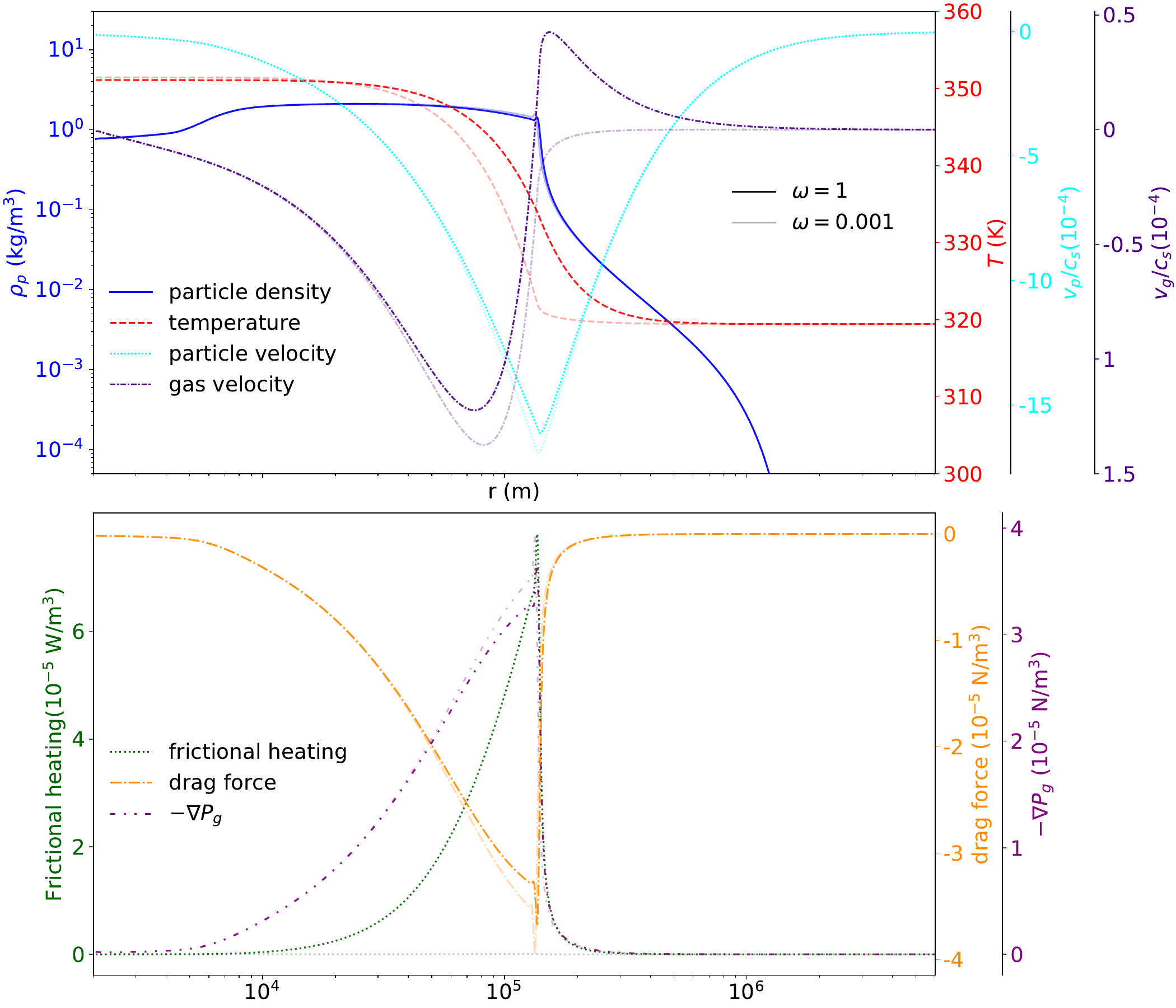}
 \caption{Radial profiles of (top) pebble density (blue, left), gas temperature (red, right), pebble velocity (cyan, right), gas velocity (indigo, right), and (bottom) frictional heating (green, left), drag force (orange, right) and gas pressure gradient (violet, right) at $t/t_{ff}=4.756$ for $\omega=1$ (fiducial case) and $\omega=0.001$}
  \label{final_profiles}
\end{figure}

We now consider particle drag as a precursory process initiating the gas compression responsible for the temperature increase. We compare the gas temperature with the adiabatic evolution given by the Laplace law for perfect gas (dashed curve in Figure \ref{compar_om}). The two curves perfectly overlap, showing that adiabatic heating is the main process responsible for the gas temperature increase at the centre of the cloud. To identify how the adiabatic compression occurs, we study in Figure \ref{final_profiles} the radial profiles computed for the two extreme values of $\omega$ at the same run time. One can see that the density radial profiles (blue), and the central and outer temperature (red) perfectly match in the two runs. This is consistent with the irreversible term being negligible in the collapse dynamics. A temperature mismatch is seen only at the cloud rim that coincides with the region of non-negligible frictional heating ($\omega=1$). The irreversible term of equation \ref{edrag} has a non-negligible contribution only at the edge of the pebbles cloud. In this region, the temperature increase is higher for $\omega=1$ than $\omega=0.001$. Finally, the amplitude of the (negative) drag force that does not vary with $\omega$, is maximum also at the cloud rim. Indeed, most of the pebbles' momentum is at the cloud edge, so this layer dominates the work of the drag force transferred to the gas. Furthermore, one can see that the gas pressure gradient and the drag force are nearly opposite. Consequently, the pebble drag action on the gas can be conceptualized as a piston compressing and heating the inner gas in an adiabatic process.

\subsection{Thermal exchanges between gas and pebbles.}

In this section, we consider the thermal exchanges between the gas and the pebbles that were previously neglected. To include these thermal exchanges, we now also consider the time evolution of the specific internal energy of the pebbles $u_d$. It is advected with the flow of pebbles with velocity $v_d$ and solves
\begin{equation}
    \partial_t(\rho_d u_d)+\nabla\cdot(v_d\rho_d u_d) = S_d
\end{equation}
$S_d$ is the source term modelling representing the thermal energy transfer from the gas to the pebbles. It is therefore subtracted from the gas energy equation to ensure the energy conservation of the whole system.

A rigorous estimation of these thermal exchanges is complex and would need a precise understanding of the structure of the pebbles (size, shape, composition, porosity,...). We choose here to set $S_d$ as a heat transfer at the surface of spherical pebbles and write \citep[see e.g.][equation 14.1.1]{bird_transport_2006} 
\begin{equation}
    S_d=hs(T_g-T_d)
\end{equation}
In this equation, $h$ is the heat transfer coefficient associated with the fluid and $s$ is the exchange surface between gas and pebbles per fluid volume. $h$ is computed from the gas thermal conductivity $k$ (given by the kinetic theory of gases) and the size of pebbles $a$ through the Nusselt number $\mathrm{Nu}=\dfrac{ha}{k}$. We use the expression of the Nusselt number given by \citet[equation 14.4.5]{bird_transport_2006} $\mathrm{Nu} = 2+0.6\mathrm{Re}^{1/2}\mathrm{Pr}^{1/3}$. For the flow considered in our simulation of a perfect gas around pebbles we have, $\mathrm{Re}\ll 1$ and $\mathrm{Pr}\sim 1$, so we will consider $\mathrm{Nu} \approx 2$ and therefore $h=\dfrac{2k}{a}$. Moreover, s is given by $s=\dfrac{\rho_d}{4\pi \rho_m a^3/3}\times 4\pi a^2$ where the first term is the number density of pebbles and the second the surface of a pebble. Finally, $T_g$ is the gas temperature and $T_d$ is the temperature of the pebbles. The pebble temperature is related to the internal energy through the specific heat capacity of pebbles $C_d$. We consider the initial pebble temperature to be at the equilibrium with the gas and the internal energy is defined as the variation from the initial one. We therefore have 
\begin{equation}
    C_d(T_d-\underbrace{T_d(t=0)}_{320~\mathrm{K}})=u_d-\underbrace{u_d(t=0)}_{0}
\end{equation}
Three values of $C_d$ were considered to represent the efficiency of heat transfer through a pebble: 

\begin{itemize}
    \item $C_d$ is given by the specific heat capacity of silicates $C_{\ce{Si}}$ ($\sim \unit{900}{\joule\usk\reciprocal\kilogram\usk\reciprocal\kelvin}$). This means that the thermal diffusion inside a pebble is very efficient, and the temperature is uniform throughout the solid.
    \item $C_d=0.01C_{\ce{Si}}$: thermal diffusion is weakly efficient on timescales involved here and only 1\% of pebbles heat.
    \item $C_d=0.001C_{\ce{Si}}$: thermal diffusion is poorly efficient and only the very surface of pebbles heats.
\end{itemize}
We also note that $C_d=0$ corresponds to our previous simulations (no heating of the pebbles).
With this approach, we obtain that the resulting surface temperature of the pebbles is very similar to the gas one. The increase in gas temperature at the inner edge of the simulation strongly varies with the specific heat capacity as shown in table \ref{tab:final_temperature}. 
This is an expected result as the energy converted from the pebble velocity to internal energy is distributed into the gas and a mass of solids which strongly varies. Indeed at the end of the simulation, way before the end of the collapse, the pebble-to-gas ratio is already very high (of the order of $10^5-10^6$ as can be seen in figure \ref{profiles}) and the internal energy lies mostly in the pebbles. \\ This result questions how the total amount of internal energy that will be distributed between gas and solids, varies with the properties of the initial cloud. The thermal diffusion properties of the pebbles being unknown, we consider in the following this global internal energy of the system as represented by the gas temperature when the pebbles do not heat ($C_d=0$).

\begin{table}
    \centering
    \begin{tabular}{cc}
        case & $\Delta T_g$ \\
        \hline
        $C_d=0$ &  \unit{175}{\kelvin}\\
        $C_d=0.001C_{\ce{Si}}$ & \unit{10.7}{\kelvin} \\
        $C_d=0.01C_{\ce{Si}}$ & \unit{1.6}{\kelvin} \\
        $C_d=C_{\ce{Si}}$ & \unit{0.02}{\kelvin} \\
    \end{tabular}
    \caption{Gas temperature increase depending on the strength of thermal exchanges between gas and pebbles.}
    \label{tab:final_temperature}
\end{table}

\subsection{Parameter exploration}

\subsubsection{Initial mass and density}

\begin{figure}
 \centering
 \includegraphics[width=\columnwidth,clip]{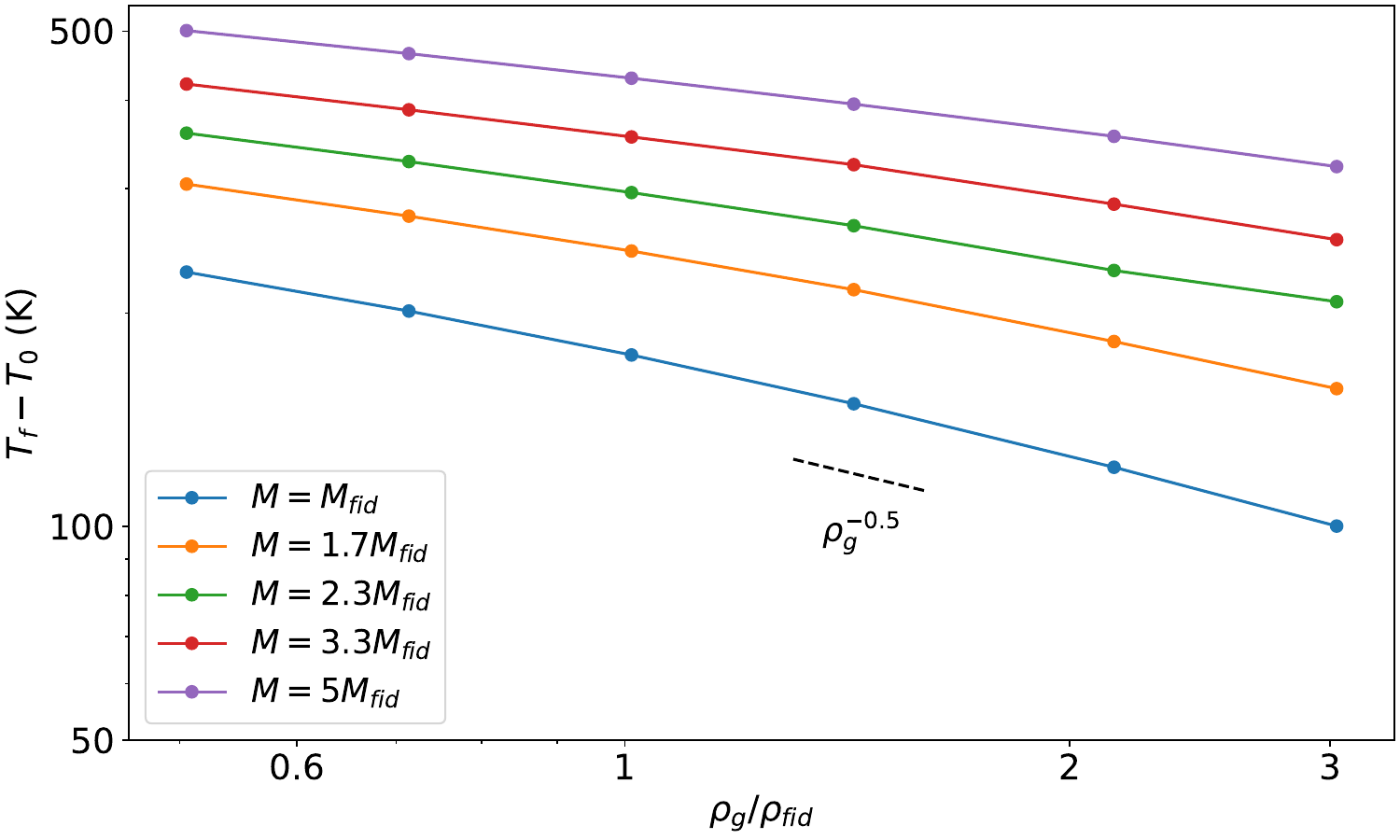}
 \caption{Increase in gas central temperature as a function of the initial gas density, for different values of the initial mass of pebbles.}
  \label{resume_rho}
\end{figure}

\begin{figure}
 \centering
 \includegraphics[width=\columnwidth,clip]{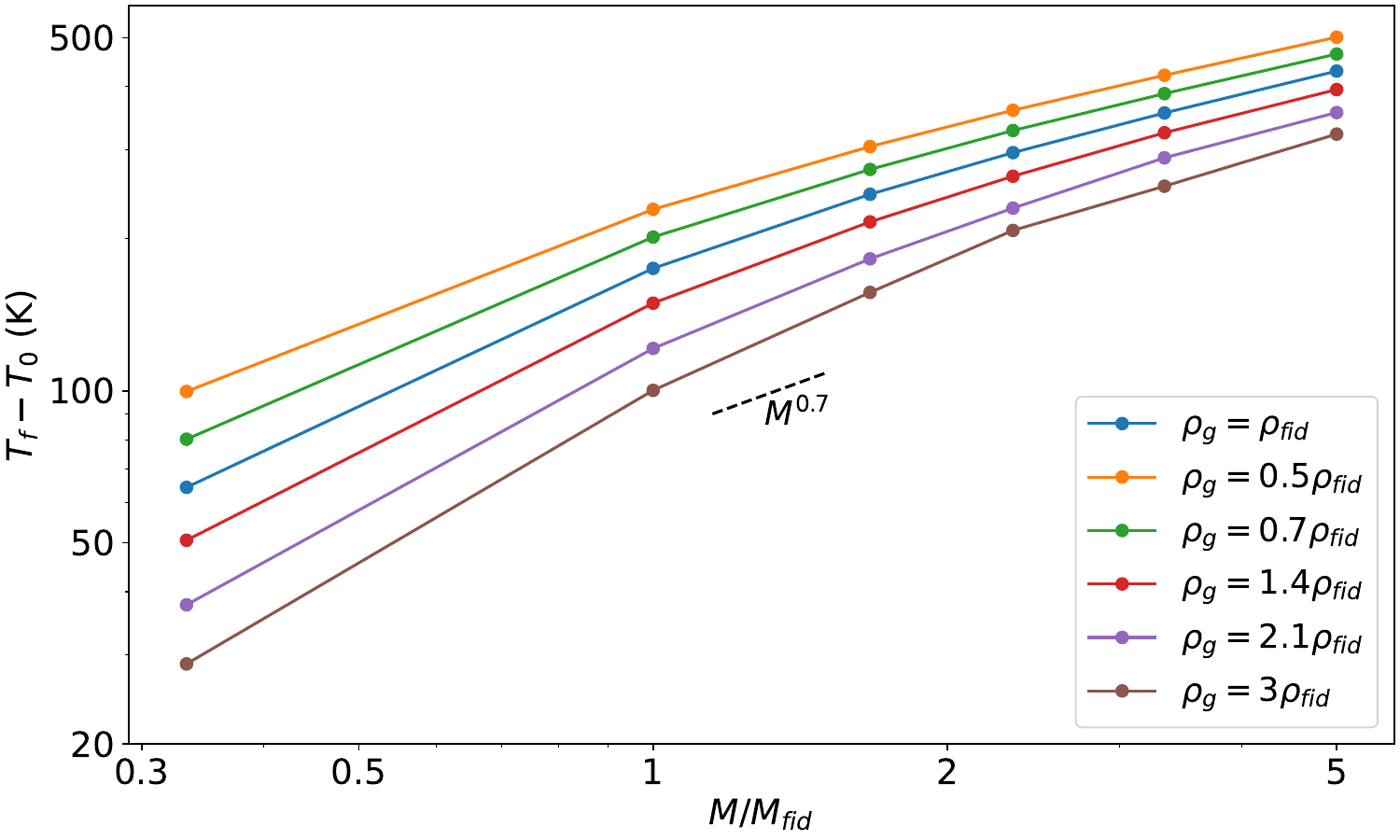}
 \caption{Increase in the gas central temperature as a function of the initial mass of pebbles for different values of initial gas density.}
  \label{resume_mass}
\end{figure}

We study how the final central gas temperature evolves with two parameters: the total mass of the pebble cloud $M_{tot}$ and the initial gas density $\rho_g$, characterized by $M_\mathrm{fid}=\unit{3\times\power{10}{16}}{\kilogram}$ and $\rho_\mathrm{fid}=\unit{7\times\power{10}{-6}}{\kilogram\usk\meter\rpcubed}$. As shown in Fig. \ref{resume_rho}, the final central temperature is approximately inversely proportional to the square root of the initial gas density. However, the slope of the curve varies slightly with the initial cloud mass. To better characterize this behaviour, the final temperature is plotted as a function of the initial mass in Fig. \ref{resume_mass}. The final temperature is nearly proportional to the initial mass, but again, with a slight dependence on the gas density.

These behaviours are consistent with what we might expect: the more massive the pebble clump, the more energy is released, and the higher the temperature of the gas surrounding the forming planetesimal. Similarly, if there is less gas surrounding the clump, the gas will be hotter, for the same amount of energy released. These results show that, to dissolve gas from the disc into the forming planetesimal (i.e. to reach a temperature of about \unit{1200}{\kelvin}), a massive clump (mass of about \unit{\power{10}{17}}{\kilogram}, corresponding to a planetesimal of about \unit{20}{\kilo\meter} in size) with a low surrounding gas density (initial dust-to-gas ratio of about 1000 or more) is required, as well as a nearly-zero thermal diffusion in the pebbles.

\subsubsection{Size and intrinsic density of pebbles}

\begin{figure}
 \centering
 \includegraphics[width=\columnwidth,clip]{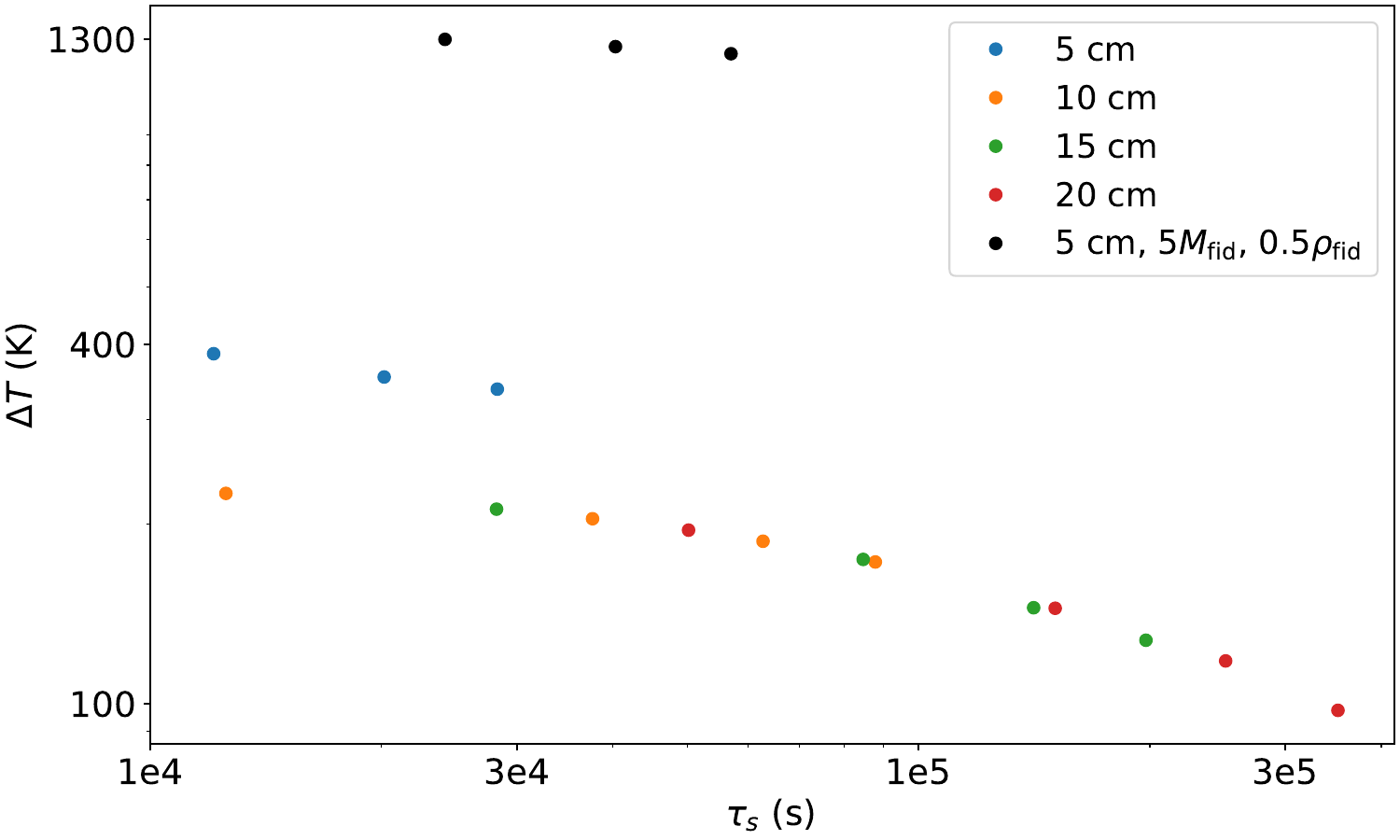}
 \caption{Gas central temperature increase as a function of the initial stopping time of pebbles.
 }
  \label{resume_taus}
\end{figure}

We also study the influence of the coupling between gas and particle by varying the initial stopping time of the pebbles. This was made by modifying either the size or the internal density of the pebbles. Sizes were chosen between 5 and \unit{20}{\centi\meter} and densities between 1 and \unit{7}{\gram\usk\centi\meter\rpcubed} (pebbles made of different materials from ice to iron depending on porosity). Because of the transition in the formulation of the stopping time between the Epstein and Stokes drag law for $s=\unit{6.9}{\centi\meter}$ in our fiducial setup where the mean free path of the gas is $\lambda=\unit{2.9}{\centi\meter}$, there is a dichotomy in behaviour between the \unit{5}{\centi\meter} pebbles and the others. If we first consider the solids larger than \unit{10}{\centi\meter}, one can see in figure \ref{resume_taus} that the smaller the initial stopping time, the hotter the gas at the end of the run. Indeed, the drag term in equations \ref{Euler_u} and \ref{Euler_e} is inversely proportional to the stopping time. Thus, for smaller and less dense pebbles, more coupled to the gas, the piston effect is more efficient and the gas temperature is higher. This correlation between the stopping time and adiabatic heating is also responsible for the dichotomy seen in Fig. \ref{resume_taus} between the 5cm size pebbles and the other, for the same initial stopping time. Indeed in a perfect gas, the mean free path $\lambda$ is proportional to $\rho_g^{-1}$. Therefore, with eq. \ref{def_tau}, we have $\tau_s \propto (\rho_g \overline{c})^{-1}$ for Epstein law and $\tau_s \propto \overline{c}^{-1}$ for Stokes law. It implies that for a fixed-size solid the modification of its stopping time with the gas density and temperature is larger in the Epstein regime. Thus, for two pebbles with the same initial stopping time, the one in the Epstein regime will become more coupled to the gas as the collapse progresses than the one in the Stokes regime. So, at the end of the simulation, the piston effect will be more efficient for small pebbles than for big ones.

Figures \ref{resume_rho}, \ref{resume_mass} and \ref{resume_taus} demonstrate that the maximum temperature increase in the gas is achieved when the initial gas density is the lowest, the total mass of the cloud is the largest or the pebbles are the smallest. In the parameter range under consideration (see Table \ref{tab:values}), the greatest temperature increase is therefore observed for $M_{tot}=5M_\mathrm{fid}$, $\rho_{g}=0.5\rho_\mathrm{fid}$ and $a=\unit{5}{\centi\meter}$. We run the corresponding simulations, whose result corresponds to the black dots in figure \ref{resume_taus}. In this case, we obtain an increase of about \unit{1300}{\kelvin} resulting in a final temperature of approximately \unit{1600}{\kelvin}.

\section{Discussion and Conclusion}\label{discussion}

To ascertain the impact of numerical resolution on the results, additional simulations were conducted using the same initial conditions as the fiducial case but with an increasing number of cells. The findings are presented in appendix \ref{convegence}. We verify that increasing the resolution does not change the physical results, whereas decreasing it could occasionally result in minor numerical instabilities.

The previous exploration of the parameter space shows the importance of the gas heating process during planetesimal formation by gravitational collapse. However, due to the simplicity of our approach, there are several limitations.

First, the collapse is assumed to be spherically symmetric. This implies that convection in the gas, as well as rotation, are neglected. In our fiducial simulation, we estimated the Brunt–Väisälä frequency, designated as $N$. It was found that the value of $N^2$ is positive in the edge zone at a radius $r\approx \unit{\power{10}{5}}{\kilo\meter}$. However, the value of $N$ is $\approx\unit{\power{10}{-4}}{\hertz}$ resulting in a characteristic time of \unit{\power{10}{4}}{\second}. This timescale is however of the order of the temperature evolution timescale, casting doubts on the viability of genuine convection. So, the temperature profiles shown in Figure \ref{profiles} reached at the end of the simulation might be prone to convection. However, this estimation is made without any cooling. This process would imply a redistribution of energy within the gas, which would reduce the final central temperature.

Finally, we do not consider particle collisions, despite their potentially key role in planetesimal formation \citep{nesvorny_formation_2010,nesvorny_binary_2021,jansson_role_2017,wahlberg_jansson_radially_2017,polak_high-resolution_2023} for two reasons. First, we run 1D simulations, with no initial velocity dispersion, making collisions meaningless here. Second, unlike \cite{nesvorny_formation_2010,nesvorny_binary_2021} who focus on Kuiper belt objects, we consider planetesimal formation closer to 1~AU with denser and hotter gas, leading to stronger friction, which may dominate the effect of collisions in the first phase of the collapse (see a comparison between stopping time and collision time in appendix \ref{comp_col_stop}). It is a reasonable assumption that collisions will play a crucial role in the subsequent evolution. This is due to the fact that collisions will facilitate a conversion of the kinetic energy of the pebbles into internal energy, which will in turn influence the temperature of the pebbles.

To conclude, in this paper, we have extended the \citet{shariff_spherically_2015} approach of the gravitational collapse of a gas and pebble cloud, by considering the gas thermodynamics. We first show an increase in the collapse time of the pebble cloud due to the gas friction. This increase is inversely proportional to the Stokes number $\Stokes_\mathrm{ff}=2\pi \tau_s / t_\mathrm{ff}$. Secondly, we have provided evidence that the heating of the gas at the centre of the domain is caused by gas adiabatic compression driven by the pebbles dragging the gas with them as they collapse. Three regions can be identified as associated with different heating. In the outermost region without particles, there is no temperature increase. Frictional heating, when present, heats only the particle cloud rim. The temperature rise in the core of the cloud is dominated by adiabatic compression driven by the deposition of momentum in the rim. We also show how this heating varies with the clump mass, the gas density and the characteristics of the pebbles. In the most massive clumps, more gravitational energy will be released during the collapse, and thus the pebbles will reach a higher velocity. Moreover, if these clumps are made of small pebbles, the coupling between pebbles and gas is stronger, which implies that the piston effect is more efficient. Therefore in these clumps, the gas might reach the temperature needed for dissolution into the forming planetesimal. Furthermore, the final stages of the collapse are not modelled in this study. If the collapse were to be simulated to a greater extent, a higher temperature should be reached. The temperature might be high enough (up to about \unit{1500}{\kelvin}) for the pebbles to experience thermal metamorphism, and, maybe, even a melting favouring a gas dissolution into it, or degassing of trapped gases. However, these results depend strongly on the thermal exchanges between gas and pebbles and on their composition. At such pebble scales, geochemical experiments of gas dissolution into silicates show that this process could be highly effective in a few hours \citep[see e.g.][]{jambon_solubility_1986}. This work demonstrates that the melting of the pebble surface might be obtained during the gravitational collapse over a timescale of $10^{-2}$ free-fall time (5 hours here) in favourable conditions (massive clump, small pebbles, low to nearly zero thermal diffusion). Therefore even for planetesimals small compared to the characteristic size of parent bodies of the current main belt asteroids, the temperature increase might be consistent with a gas dissolution in the planetesimal-forming pebbles. However, one should note that the cooling process after the planetesimal formation is not considered here.

These results emphasize the key role of the gas in the formation of planetesimals with gravitational collapse, especially at distances of a few AUs from the central star. This implies a new step in the thermal history of planetesimals depending strongly on their formation region. 
The identified heating process may lead to local heating not only in space but also in time which could provide new approaches for cosmochemical studies. This work also suggests future avenues for studying the gravitational collapse of a pebble clump in a gaseous environment: first, by extending the simulations to 2D or 3D, opening up the possibility of including initial particle velocity dispersion and collisions, and second, by better modelling the thermal exchange between pebbles and gas.

\begin{acknowledgements}

This work was supported by the French government through the France 2030 investment plan managed by the National Research Agency (ANR), as part of the Initiative of Excellence Université Côte d’Azur under reference number ANR-15-IDEX-01. The authors are grateful to the Université Côte d’Azur’s Center for High-Performance Computing (OPAL infrastructure) for providing resources and support. It has received support from the CNRS Origines and 'Programme national de Planétologie' (PNP). The authors would like to thank the anonymous referee for the very useful comments that helped to improve greatly this paper.

\end{acknowledgements}

\bibliographystyle{aa}
\bibliography{biblio}

\begin{thebibliography}{48}
\expandafter\ifx\csname natexlab\endcsname\relax\def\natexlab#1{#1}\fi

\bibitem[{Bird {et~al.}(2006)Bird, Stewart, \& Lightfoot}]{bird_transport_2006}
Bird, R., Stewart, W., \& Lightfoot, E. 2006, Transport {Phenomena} (Wiley)

\bibitem[{Blum(2018)}]{blum_dust_2018}
Blum, J. 2018, Space Sci. Rev., 214, 52

\bibitem[{Boyet {et~al.}(2018)Boyet, Bouvier, Frossard, Hammouda, Garçon, \&
  Gannoun}]{boyet_enstatite_2018}
Boyet, M., Bouvier, A., Frossard, P., {et~al.} 2018, Earth Planet Sci. Lett.,
  488, 68

\bibitem[{Carrera \& Simon(2022)}]{carrera_streaming_2022}
Carrera, D. \& Simon, J.~B. 2022, ApJL, 933, L10

\bibitem[{Cuzzi {et~al.}(2001)Cuzzi, Hogan, Paque, \&
  Dobrovolskis}]{cuzzi_size-selective_2001}
Cuzzi, J.~N., Hogan, R.~C., Paque, J.~M., \& Dobrovolskis, A.~R. 2001, ApJ,
  546, 496

\bibitem[{Dittrich {et~al.}(2013)Dittrich, Klahr, \&
  Johansen}]{dittrich_gravoturbulent_2013}
Dittrich, K., Klahr, H., \& Johansen, A. 2013, ApJ, 763, 117

\bibitem[{Epstein(1924)}]{epstein_resistance_1924}
Epstein, P.~S. 1924, Phys. Rev., 23, 710

\bibitem[{Fromang \& Nelson(2006)}]{fromang_global_2006}
Fromang, S. \& Nelson, R.~P. 2006, A\&A, 457, 343

\bibitem[{Gerosa {et~al.}(2023)Gerosa, Méheut, \& Bec}]{gerosa_clusters_2023}
Gerosa, F.~A., Méheut, H., \& Bec, J. 2023, Eur. Phys. J. Plus, 138, 9

\bibitem[{Gratton {et~al.}(2019)Gratton, Ligi, Sissa, Desidera, Mesa, Bonnefoy,
  Chauvin, Cheetham, Feldt, Lagrange, Langlois, Meyer, Vigan, Boccaletti,
  Janson, Lazzoni, Zurlo, De~Boer, Henning, D’Orazi, Gluck, Madec, Jaquet,
  Baudoz, Fantinel, Pavlov, \& Wildi}]{gratton_blobs_2019}
Gratton, R., Ligi, R., Sissa, E., {et~al.} 2019, A\&A, 623, A140

\bibitem[{Hasegawa {et~al.}(2021)Hasegawa, Suzuki, Tanaka, Kobayashi, \&
  Wada}]{hasegawa_collisional_2021}
Hasegawa, Y., Suzuki, T.~K., Tanaka, H., Kobayashi, H., \& Wada, K. 2021, ApJ,
  915, 22

\bibitem[{Hasegawa {et~al.}(2023)Hasegawa, Suzuki, Tanaka, Kobayashi, \&
  Wada}]{hasegawa_collisional_2023}
Hasegawa, Y., Suzuki, T.~K., Tanaka, H., Kobayashi, H., \& Wada, K. 2023, ApJ,
  944, 38

\bibitem[{Hayashi(1981)}]{hayashi_structure_1981}
Hayashi, C. 1981, Prog. Theor. Phys. Supp., 70, 35

\bibitem[{Huang \& Bai(2022)}]{huang_multifluid_2022}
Huang, P. \& Bai, X.-N. 2022, \apjs, 262, 11

\bibitem[{Jambon {et~al.}(1986)Jambon, Weber, \&
  Braun}]{jambon_solubility_1986}
Jambon, A., Weber, H., \& Braun, O. 1986, Geochim. Cosmochim. Acta, 50, 401

\bibitem[{Johansen {et~al.}(2011)Johansen, Klahr, \&
  Henning}]{johansen_high-resolution_2011}
Johansen, A., Klahr, H., \& Henning, T. 2011, A\&A, 529, A62

\bibitem[{Johansen \& Lambrechts(2017)}]{johansen_forming_2017}
Johansen, A. \& Lambrechts, M. 2017, Annu. Rev. Earth Planet. Sci., 45, 359

\bibitem[{Keppler {et~al.}(2018)Keppler, Benisty, Müller, Henning, Van~Boekel,
  Cantalloube, Ginski, Van~Holstein, Maire, Pohl, Samland, Avenhaus, Baudino,
  Boccaletti, De~Boer, Bonnefoy, Chauvin, Desidera, Langlois, Lazzoni, Marleau,
  Mordasini, Pawellek, Stolker, Vigan, Zurlo, Birnstiel, Brandner, Feldt,
  Flock, Girard, Gratton, Hagelberg, Isella, Janson, Juhasz, Kemmer, Kral,
  Lagrange, Launhardt, Matter, Ménard, Milli, Mollière, Olofsson, Pérez,
  Pinilla, Pinte, Quanz, Schmidt, Udry, Wahhaj, Williams, Buenzli, Cudel,
  Dominik, Galicher, Kasper, Lannier, Mesa, Mouillet, Peretti, Perrot, Salter,
  Sissa, Wildi, Abe, Antichi, Augereau, Baruffolo, Baudoz, Bazzon, Beuzit,
  Blanchard, Brems, Buey, De~Caprio, Carbillet, Carle, Cascone, Cheetham,
  Claudi, Costille, Delboulbé, Dohlen, Fantinel, Feautrier, Fusco, Giro,
  Gluck, Gry, Hubin, Hugot, Jaquet, Le~Mignant, Llored, Madec, Magnard,
  Martinez, Maurel, Meyer, Möller-Nilsson, Moulin, Mugnier, Origné, Pavlov,
  Perret, Petit, Pragt, Puget, Rabou, Ramos, Rigal, Rochat, Roelfsema, Rousset,
  Roux, Salasnich, Sauvage, Sevin, Soenke, Stadler, Suarez, Turatto, \&
  Weber}]{keppler_discovery_2018}
Keppler, M., Benisty, M., Müller, A., {et~al.} 2018, A\&A, 617, A44

\bibitem[{Klahr \& Hubbard(2014)}]{klahr_convective_2014}
Klahr, H. \& Hubbard, A. 2014, ApJ, 788, 21

\bibitem[{Kobayashi \& Tanaka(2021)}]{kobayashi_rapid_2021}
Kobayashi, H. \& Tanaka, H. 2021, ApJ, 922, 16

\bibitem[{Kobayashi \& Tanaka(2023)}]{kobayashi_rapid_2023}
Kobayashi, H. \& Tanaka, H. 2023, ApJ, 954, 158

\bibitem[{Krapp {et~al.}(2019)Krapp, Benítez-Llambay, Gressel, \&
  Pessah}]{krapp_streaming_2019}
Krapp, L., Benítez-Llambay, P., Gressel, O., \& Pessah, M.~E. 2019, ApJL, 878,
  L30

\bibitem[{Kruijer {et~al.}(2014)Kruijer, Touboul, Fischer-Gödde, Bermingham,
  Walker, \& Kleine}]{kruijer_protracted_2014}
Kruijer, T.~S., Touboul, M., Fischer-Gödde, M., {et~al.} 2014, Science, 344,
  1150

\bibitem[{{Lesur} {et~al.}(2023){Lesur}, {Flock}, {Ercolano}, {Lin}, {Yang},
  {Barranco}, {Benitez-Llambay}, {Goodman}, {Johansen}, {Klahr}, {Laibe},
  {Lyra}, {Marcus}, {Nelson}, {Squire}, {Simon}, {Turner}, {Umurhan}, \&
  {Youdin}}]{lesur_hydro-_2022}
{Lesur}, G., {Flock}, M., {Ercolano}, B., {et~al.} 2023, in Astronomical
  Society of the Pacific Conference Series, Vol. 534, Protostars and Planets
  VII, ed. S.~{Inutsuka}, Y.~{Aikawa}, T.~{Muto}, K.~{Tomida}, \& M.~{Tamura},
  465

\bibitem[{Lesur {et~al.}(2023)Lesur, Baghdadi, Wafflard-Fernandez, Mauxion,
  Robert, \& Van Den~Bossche}]{lesur_idefix_2023}
Lesur, G. R.~J., Baghdadi, S., Wafflard-Fernandez, G., {et~al.} 2023, A\&A,
  677, A9

\bibitem[{{Lichtenberg} {et~al.}(2023){Lichtenberg}, {Schaefer}, {Nakajima}, \&
  {Fischer}}]{lichtenberg_geophysical_2022}
{Lichtenberg}, T., {Schaefer}, L.~K., {Nakajima}, M., \& {Fischer}, R.~A. 2023,
  in Astronomical Society of the Pacific Conference Series, Vol. 534,
  Protostars and Planets VII, ed. S.~{Inutsuka}, Y.~{Aikawa}, T.~{Muto},
  K.~{Tomida}, \& M.~{Tamura}, 907

\bibitem[{Lorek \& Johansen(2024)}]{lorek_formation_2024}
Lorek, S. \& Johansen, A. 2024, A\&A, 683, A38

\bibitem[{Lovelace {et~al.}(1999)Lovelace, Li, Colgate, \&
  Nelson}]{lovelace_rossby_1999}
Lovelace, R. V.~E., Li, H., Colgate, S.~A., \& Nelson, A.~F. 1999, ApJ, 513,
  805

\bibitem[{Lyra(2014)}]{lyra_convective_2014}
Lyra, W. 2014, ApJ, 789, 77

\bibitem[{Manger \& Klahr(2018)}]{manger_vortex_2018}
Manger, N. \& Klahr, H. 2018, MNRAS, 480, 2125

\bibitem[{{Mauxion} {et~al.}(2024){Mauxion}, {Lesur}, \&
  {Maret}}]{mauxion_modeling_2024}
{Mauxion}, J., {Lesur}, G., \& {Maret}, S. 2024, \aap, 686, A253

\bibitem[{Meheut {et~al.}(2012)Meheut, Meliani, Varniere, \&
  Benz}]{meheut_dust-trapping_2012}
Meheut, H., Meliani, Z., Varniere, P., \& Benz, W. 2012, A\&A, 545, A134

\bibitem[{Mignone {et~al.}(2018)Mignone, Bodo, Vaidya, \&
  Mattia}]{mignone_particle_2018}
Mignone, A., Bodo, G., Vaidya, B., \& Mattia, G. 2018, ApJ, 859, 13

\bibitem[{Morbidelli \& Nesvorný(2020)}]{morbidelli_chapter_2020}
Morbidelli, A. \& Nesvorný, D. 2020, in The {Trans}-{Neptunian} {Solar}
  {System}, ed. D.~Prialnik, M.~A. Barucci, \& L.~A. Young (Elsevier), 25--59

\bibitem[{Nelson {et~al.}(2013)Nelson, Gressel, \&
  Umurhan}]{nelson_linear_2013}
Nelson, R.~P., Gressel, O., \& Umurhan, O.~M. 2013, MNRAS, 435, 2610

\bibitem[{Nesvorný {et~al.}(2021)Nesvorný, Li, Simon, Youdin, Richardson,
  Marschall, \& Grundy}]{nesvorny_binary_2021}
Nesvorný, D., Li, R., Simon, J.~B., {et~al.} 2021, Planet. Sci. J., 2, 27

\bibitem[{Nesvorný {et~al.}(2010)Nesvorný, Youdin, \&
  Richardson}]{nesvorny_formation_2010}
Nesvorný, D., Youdin, A.~N., \& Richardson, D.~C. 2010, AJ, 140, 785

\bibitem[{Pan {et~al.}(2011)Pan, Padoan, Scalo, Kritsuk, \&
  Norman}]{pan_turbulent_2011}
Pan, L., Padoan, P., Scalo, J., Kritsuk, A.~G., \& Norman, M.~L. 2011, ApJ,
  740, 6

\bibitem[{Polak \& Klahr(2023)}]{polak_high-resolution_2023}
Polak, B. \& Klahr, H. 2023, ApJ, 943, 125

\bibitem[{Raettig {et~al.}(2021)Raettig, Lyra, \& Klahr}]{raettig_pebble_2021}
Raettig, N., Lyra, W., \& Klahr, H. 2021, ApJ, 913, 92

\bibitem[{Schaffer {et~al.}(2021)Schaffer, Johansen, \&
  Lambrechts}]{schaffer_streaming_2021}
Schaffer, N., Johansen, A., \& Lambrechts, M. 2021, A\&A, 653, A14

\bibitem[{Shariff \& Cuzzi(2015)}]{shariff_spherically_2015}
Shariff, K. \& Cuzzi, J.~N. 2015, ApJ, 805, 42

\bibitem[{{Urpin} \& {Brandenburg}(1998)}]{urpin_magnetic_1998}
{Urpin}, V. \& {Brandenburg}, A. 1998, \mnras, 294, 399

\bibitem[{Wahlberg~Jansson \& Johansen(2017)}]{wahlberg_jansson_radially_2017}
Wahlberg~Jansson, K. \& Johansen, A. 2017, MNRAS, 469, S149

\bibitem[{Wahlberg~Jansson {et~al.}(2017)Wahlberg~Jansson, Johansen, Syed, \&
  Blum}]{jansson_role_2017}
Wahlberg~Jansson, K., Johansen, A., Syed, M.~B., \& Blum, J. 2017, ApJ, 835,
  109

\bibitem[{Weidenschilling(1977)}]{weidenschilling_aerodynamics_1977}
Weidenschilling, S.~J. 1977, MNRAS, 180, 57

\bibitem[{Youdin \& Goodman(2005)}]{youdin_streaming_2005}
Youdin, A.~N. \& Goodman, J. 2005, ApJ, 620, 459

\bibitem[{Zhu {et~al.}(2023)Zhu, Schiller, Moynier, Groen, Alexander, Davidson,
  Schrader, Bischoff, \& Bizzarro}]{zhu_chondrite_2023}
Zhu, K., Schiller, M., Moynier, F., {et~al.} 2023, Geochim. Cosmochim. Acta,
  342, 156

\end{thebibliography}

\begin{appendix}

\section{Derivation of analytic formulas}

\subsection{Free-fall time}\label{ff_time}

The free-fall time of a particle clump can be calculated using the same reasoning as in section \ref{times}. The position of a free-falling particle at the cloud edge solves
\begin{equation}
\dfrac{\mathrm{d}^2 r}{\mathrm{d}t^2}=-\dfrac{\mathcal{G}M}{r^2}
\label{eq_diff_free}
\end{equation}
This equation can be integrated for a static initial condition and an initial radius $R_0$ giving
\begin{equation}
\dfrac{\mathrm{d} r}{\mathrm{d}t}=-\sqrt{2\mathcal{G}M\left(\dfrac{1}{r}-\dfrac{1}{R_0}\right)}, \text{ and}
\end{equation}
\begin{equation}
t = \sqrt{\dfrac{R_0^3}{2\mathcal{G}M}}\left(\sqrt{x(1-x)}+\arcsin(\sqrt{1-x})\right),
\label{t_eff}
\end{equation}
where $x = {r}/{R_0}$. The free-fall time corresponds to the time where $r$ (or $x$) reaches 0, which gives 
\begin{equation}t_\mathrm{ff}=\dfrac{\pi}{2}\sqrt{\dfrac{R_0^3}{2\mathcal{G}M}}\end{equation}

\subsection{Comparison of collision time versus stopping time}\label{comp_col_stop}

Building on \citet{nesvorny_formation_2010}, we compare the stopping time and the time between pebble collisions. The stopping time is estimated in the Epstein regime as $\tau_s={\rho_m a}/{\rho \overline{c}}$ with the notations defined in section \ref{gov_eq}. The collision time is given by $\tau_c\sim({n\sigma v})^{-1}$, with the pebbles number density $n\sim{M_{tot}}/({\rho_m a^3 R^3})$, the cross-section $\sigma\sim a^2$, and the virial velocity $v\sim\sqrt{{\mathcal{G}M_{tot}}/{R}}$. Bringing all together leads to
\begin{equation}
    \dfrac{\tau_s}{\tau_c}\sim \sqrt{\dfrac{\mathcal{G}M_{tot}^3}{\rho^2\overline{c}^2R^7}}.
\end{equation}
With the values for $M_\mathrm{tot}$, $\rho$, $\overline{c}$ and $R$ used in our fiducial simulation, we have $\tau_s/\tau_c \sim 1$. Thus, the stopping time and the collapse time are approximately the same in this situation and thus the friction due to the gas can not be neglected. However, we have here $v\sim\unit{1}{\meter\usk\reciprocal \second}$ which is an order of magnitude higher than the typical velocity observed in our simulation (a so high velocity is obtained only at the end of the collapse). We can then consider that for our simulations, $\tau_s \ll \tau_c$ so that the friction is dominant for the major part of the collapse.

\section{Test of the numerical setup} \label{test_setup}

\subsection{Comparison with previous work}\label{comparison}

To test our numerical setup we provide here a comparison with the work of \citet{shariff_spherically_2015} where they use a similar approach for the modelisation of the system but with different numerical methods to solve hydrodynamics and Poisson equations. Moreover, we test the lagrangian particle module of \texttt{IDEFIX}, as the dynamic of the pebbles is this time solved very differently. 

\begin{figure}
 \centering
 \includegraphics[width=\columnwidth,clip]{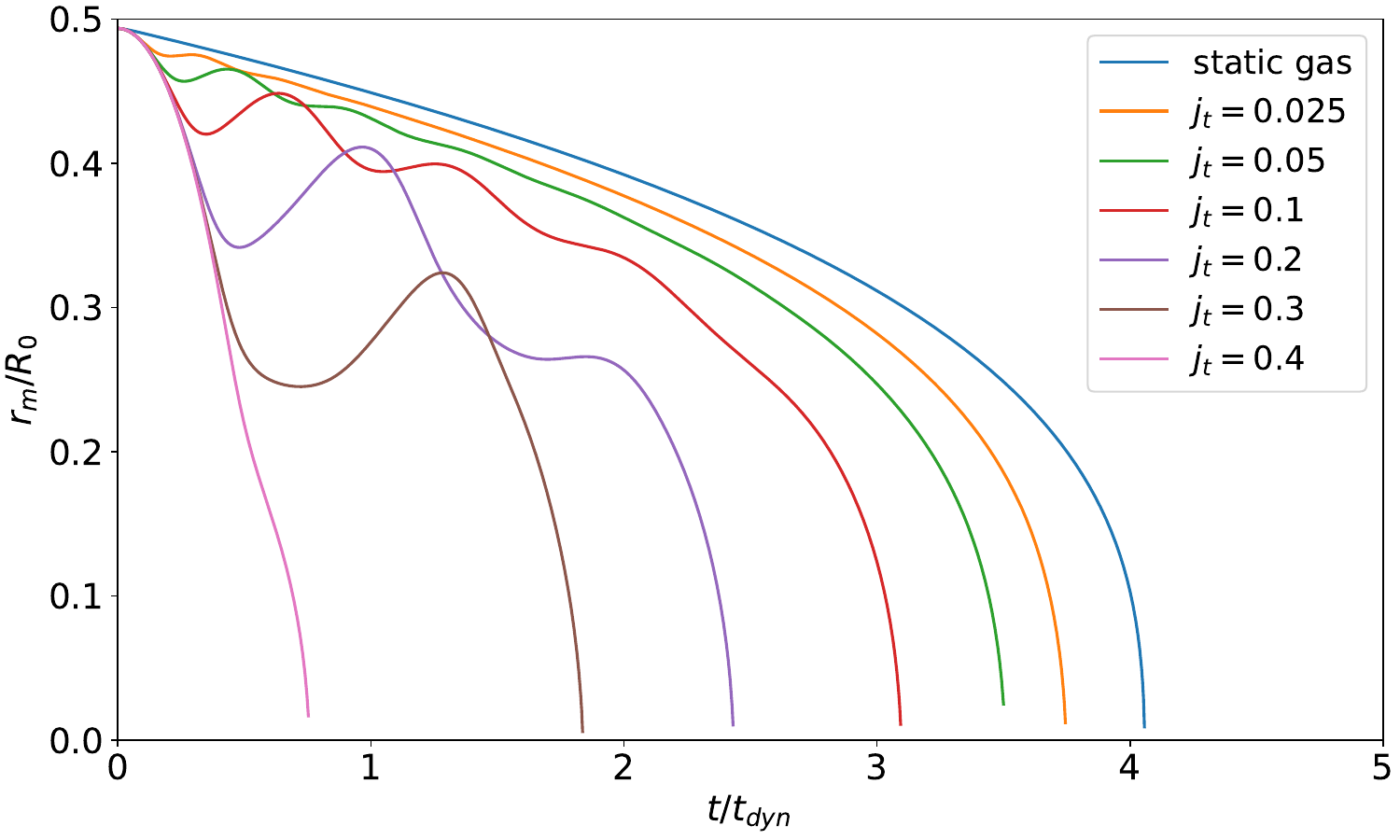}
 \caption{Size evolution of the particle cloud for various $J_t$ ($\Stokes_\mathrm{ff}=0.02$, $\phi_0=100$) with our Eulerian approach for pebbles}
  \label{compar_Shariff_multi}
\end{figure}

The notations used here are the same as section 4.1 of \cite{shariff_spherically_2015}. $J_t$ corresponds to a two-phase Jeans number which compares the propagation time of a sound wave across the clump with the dynamical time $t_{dyn}=({\rho_{p0}\mathcal{G}})^{-1/2}=\sqrt{{32}/{3\pi}}t_\mathrm{ff}$. Figure \ref{compar_Shariff_multi} compares the size evolution of the clump for various $J_t$ and should be compared with Figure 3 of \cite{shariff_spherically_2015}. We obtain results similar to the original study, which justifies our methodological approach to the problem.  

\subsection{Convergence test}\label{convegence}

\begin{figure}
 \centering
 \includegraphics[width=\columnwidth]{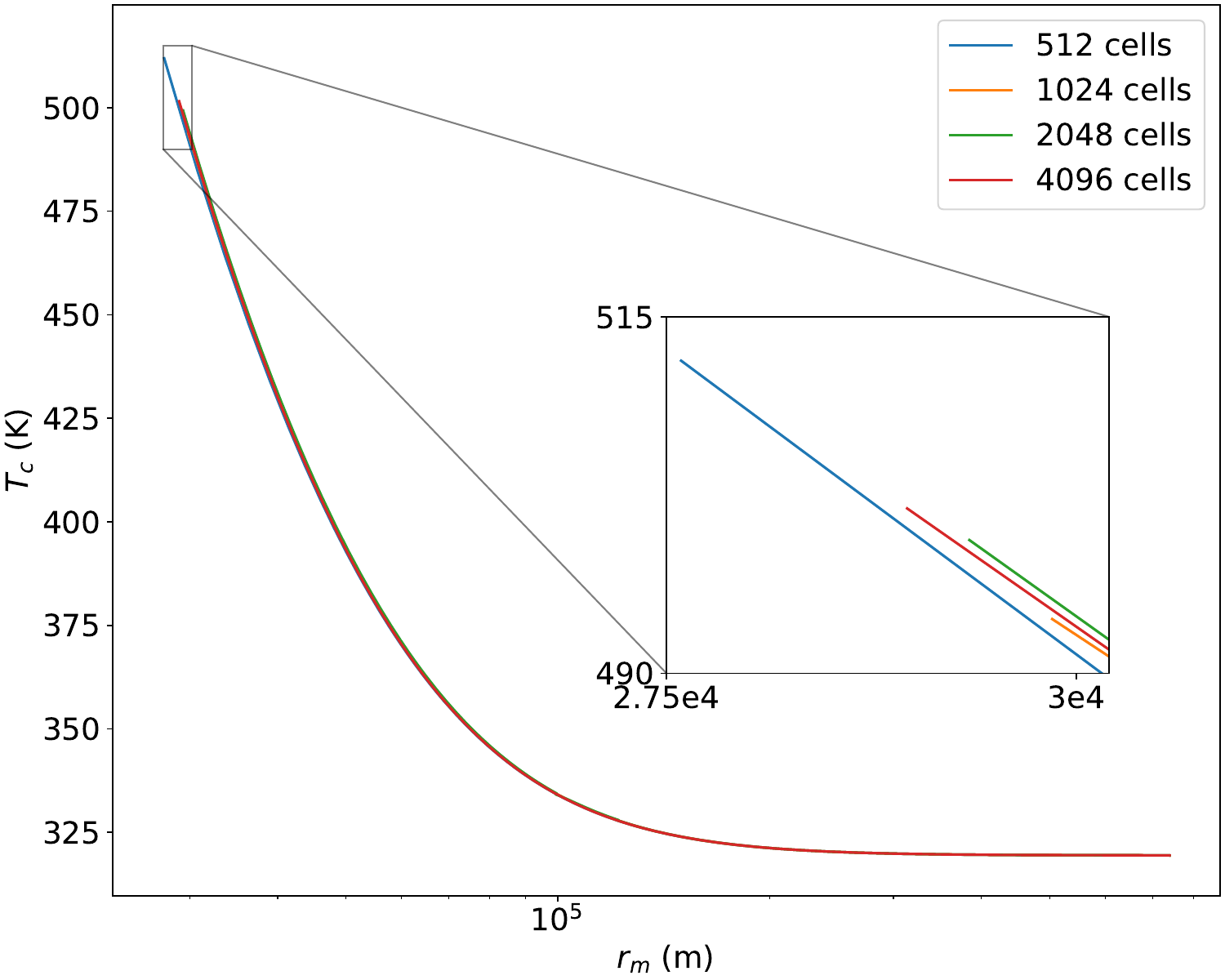}
 \caption{Evolution of the gas central temperature with the cloud mean radius for different numerical refinements.}
  \label{compar_raff}
\end{figure}

\begin{figure*}
\centering
\includegraphics[width=\textwidth]{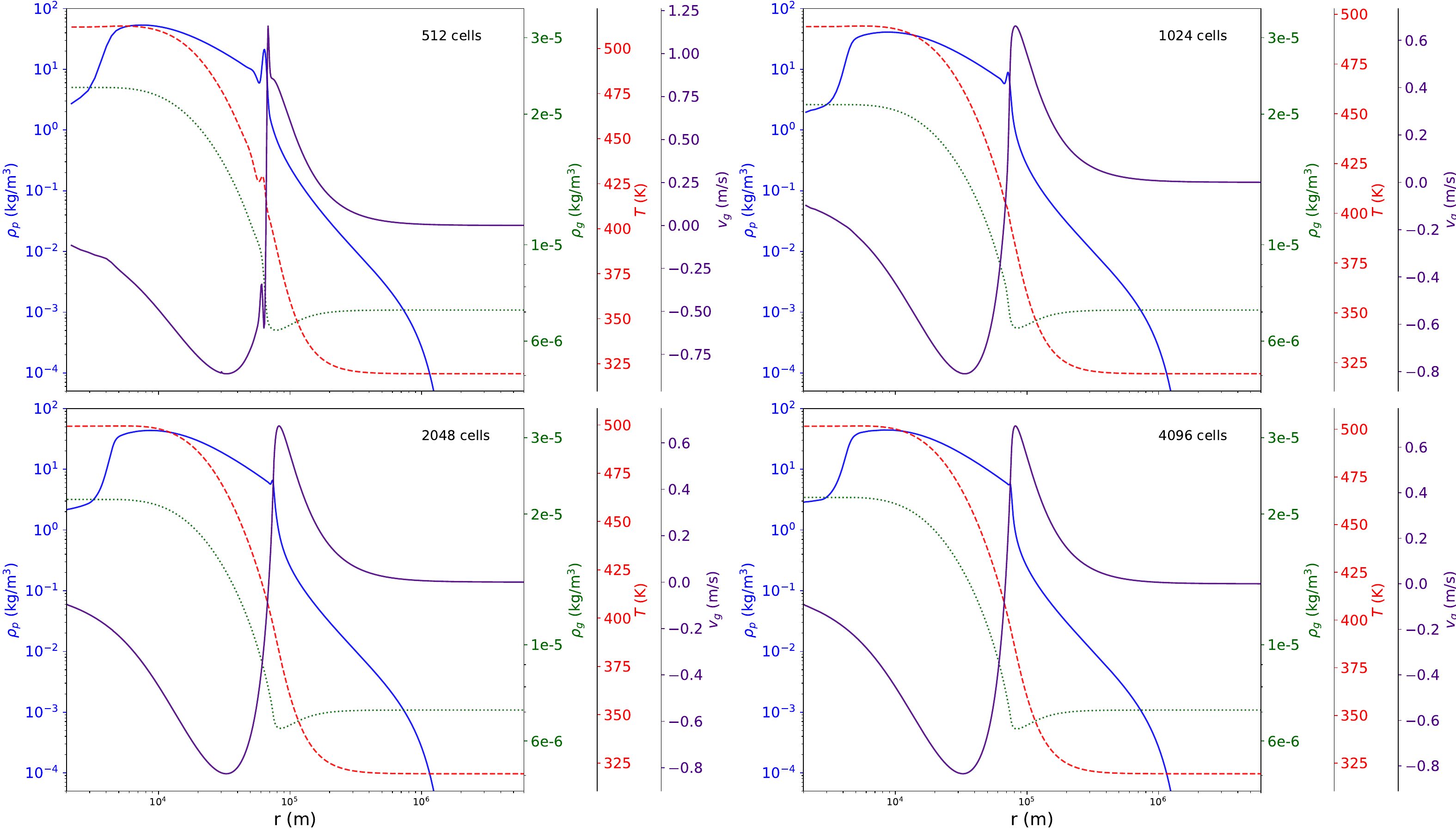}
\caption{Radial profiles of pebble density (blue), gas density (green), gas temperature (red), and gas velocity (indigo) at the end of the fiducial simulation ($t/t_{ff}=4.756$) for different numerical refinement.}

\label{compar_raff_prof}
\end{figure*}

We run our fiducial run for different resolutions, from 512 to 4096 cells. The results are shown in Figures \ref{compar_raff} and \ref{compar_raff_prof}. We see that increasing the resolution does not change the physical results (small variation of about \unit{2}{\kelvin}, which is 1\% of the temperature increase at the end of the collapse). However, we see that at low resolution there are some numerical instabilities at the edge of the cloud. We therefore decide to use a resolution of 2048 cells as a compromise between these instabilities and the simulation time.

\newpage

\section{Lagrangian Particle approach}\label{LP}

Two principal methods are typically employed to model the dynamical evolution of solids in gaseous proto-planetary discs. The first is the pressure-less fluid model, which is the focus of this paper. The second method is the Lagrangian particle approach, which involves the modelling of solids through a relatively small number of particles (referred to as super-particles, or SP). Each SP represents a large number of pebbles sharing the same properties, including size, internal density, velocity.  The \texttt{IDEFIX} code enables the utilisation of these two approaches, and thus a comparison was undertaken. The principal characteristics of the Lagrangian particles module of the \texttt{IDEFIX} code are presented here, with a more comprehensive account to follow in a dedicated publication.

The SP follow the Newton equation: 
\begin{equation}
    \dfrac{\mathrm{d}^2 \boldsymbol{x_\mathrm{p}}}{\mathrm{d}t^2} = -\boldsymbol{\nabla}(\phi_\mathrm{g+p})-\dfrac{\boldsymbol{v_\mathrm{p}} - \boldsymbol{u}}{\tau_s}
\end{equation}
where $\boldsymbol{x_\mathrm{p}}$ and $\boldsymbol{v_\mathrm{p}}$ are the position and the velocity of a SP. In order to compute the self-gravity and friction forces, the deposition of SP on the grid is carried out using a triangular shape cloud (TSC) scheme, as described by \cite{mignone_particle_2018}.

\begin{figure}
 \centering
 \includegraphics[width=\columnwidth,clip]{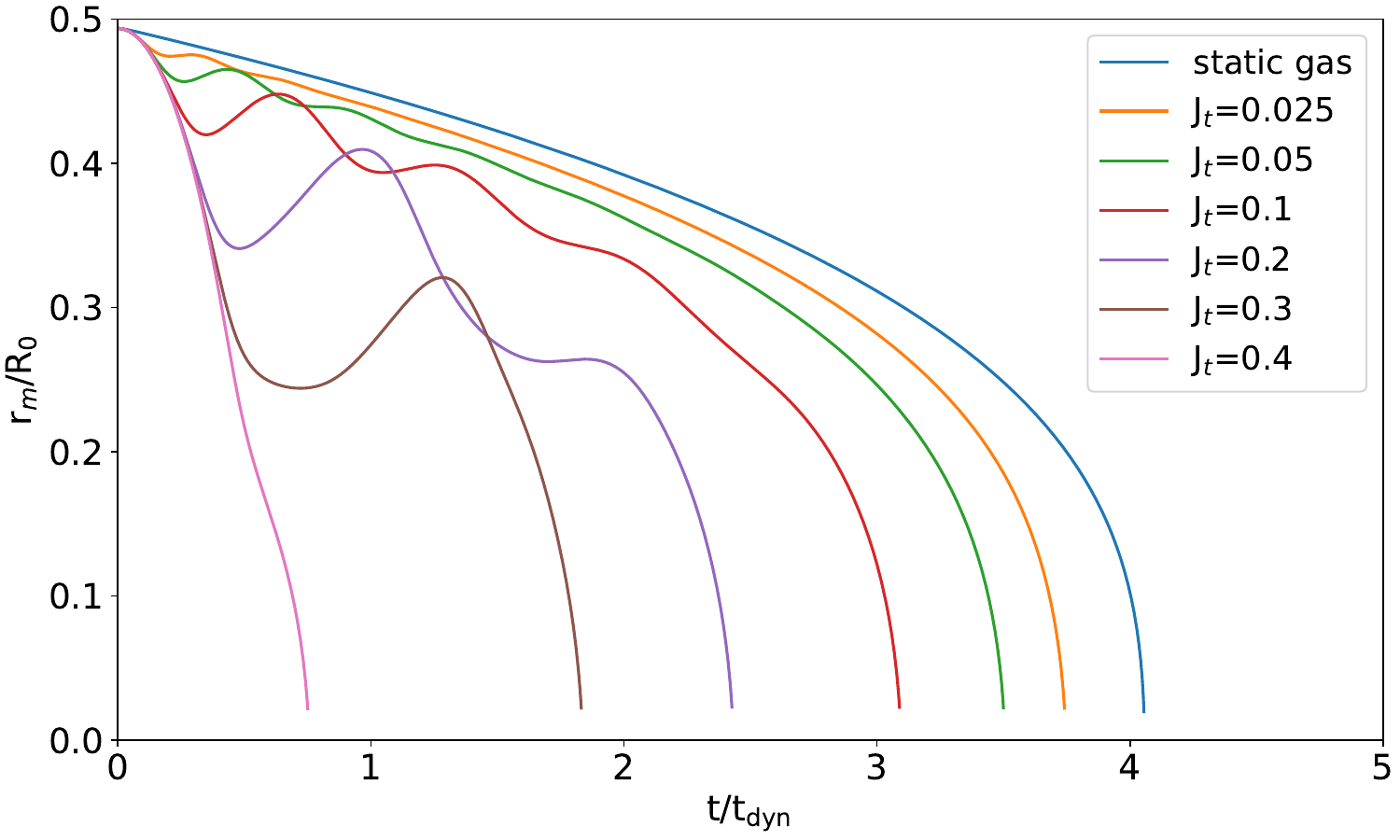}
 \caption{Size evolution of the particle cloud for various $J_t$ ($\Stokes_\mathrm{ff}=0.02$, $\phi_0=100$) with a Lagrangian approach for pebbles}
  \label{compar_Shariff}
\end{figure}

\begin{figure*}[ht!]
\centering
\includegraphics[width=\textwidth]{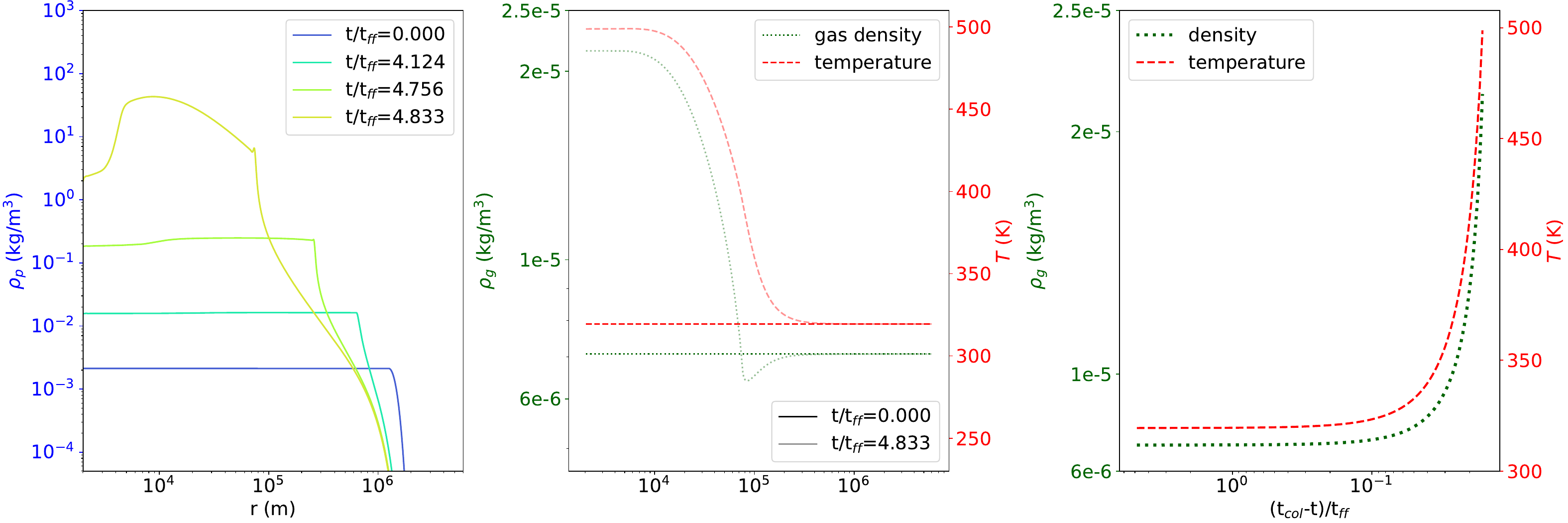}
\caption{(Left) Time evolution of the radial profile of the particle cloud density. (Centre) Gas temperature profiles (dashed) and gas density profiles (dotted), plotted at the beginning and the end of the fiducial run. (Right) Time evolution of the central density (dotted) and temperature (dashed, right axis) of the gas.}

\label{profiles_part}
\end{figure*}

The simulation of the collapse employs the identical initial conditions and grid as the pressure-less fluid approach, with a single minor alteration. To reduce the computational time, an evolving mesh refinement is implemented. The simulation is conducted in three stages, corresponding to the intervals $[0.5, 0.15]R_0$, $[0.15,0.05]R_0$, and $[0.05,0.02]R_0$. At each restart, the resolution is increased from $1024$ to $2048$ and $4096$ cells. This results in a reduction in the requirement for computational resources during the initial phase of the evolution process, while simultaneously enabling high precision during the accelerated evolution phase at the late stages of the collapse. In order to ensure continuity in cell size between the uniform zone and the logarithmic zone of the grid, the cells are distributed between the two zones in the same way as the simulations conducted with the pressure-less fluid approach. In order to achieve a smooth transition between two different steps, the final profiles from one step are used as the initial conditions for the subsequent step. These profiles, which relate to position, velocity and pressure (for the gas), are linearly interpolated on the new grid. In order to obtain sufficient resolution to correctly map the particle density onto the grid, each step is initialised with 10 particles per cell.  We verified that there were no physical differences if we did only one step with 2048 or 4096 cells. We use open boundary conditions, allowing SP to escape the grid, thereby precluding any possibility of re-entry. In a manner analogous to the pressure-less approach, the mass lost at the inner boundary is added to the central mass thereby contributing to $\phi_c$. 

Figure \ref{compar_Shariff} is the same as figure \ref{compar_Shariff_multi}, and figure \ref{profiles_part} is the same as figure \ref{profiles} but for the Lagrangian particle approach. The results obtained are highly comparable, indicating that both methodological approaches are viable for modelling the collapse. The methodological choice will therefore be contingent upon the additional physical effects one wishes to consider, such as collisions or size distributions.

\end{appendix}
\end{document}